\documentclass[11pt,a4paper]{article}
\pdfoutput=1

\usepackage{jheppub}
\usepackage{latexsym}
\usepackage{amstext}
\usepackage{amssymb}
\usepackage{amsmath}
\usepackage{multirow}
\usepackage{color}
\usepackage[usenames,dvipsnames,table]{xcolor}
\usepackage{graphicx}
\usepackage{epsfig}  
\usepackage{epsf}    
\usepackage{dcolumn}
\usepackage{bm}
\usepackage{dcolumn}
\usepackage{textcomp}
\usepackage{float}
\usepackage{hypcap}
\usepackage[]{hyperref}
\usepackage{makecell}
\usepackage{pifont}
\usepackage{appendix}

\usepackage[normalem]{ulem} 

\usepackage{lineno}

\usepackage{color}

\newcommand{\magenta}[1]{{\color{magenta} #1}}

\newcommand{\capdef}{}
\newcommand{\mycaption}[2][\capdef]{\renewcommand{\capdef}{#2}
       \caption[#1]{{\footnotesize #2}}}
\makeatletter
\renewcommand{\fnum@table}{\textbf{\tablename~\thetable}}
\renewcommand{\fnum@figure}{\textbf{\figurename~\thefigure}}
\makeatother

\preprint{IP/BBSR/2019-2}

\title{Constraints on Flavor-Diagonal Non-Standard Neutrino Interactions from Borexino Phase-II}

\collaboration{The Borexino Collaboration}

\author[1,2,3]{S. K. Agarwalla,}
\author[4]{M.~Agostini,}
\author[4]{K.~Altenm\"{u}ller,}
\author[4]{S.~Appel,}
\author[5]{V.~Atroshchenko,}
\author[6]{Z.~Bagdasarian,}
\author[7]{D.~Basilico,}
\author[7]{G.~Bellini,}
\author[8]{J.~Benziger,}
\author[9]{D.~Bick,}
% ...

\affiliation[1]{Institute of Physics, Sachivalaya Marg, Sainik School Post, Bhubaneswar 751005, India}
\affiliation[2]{Homi Bhabha National Institute, Training School Complex, Anushakti Nagar, Mumbai 400085, India}
\affiliation[3]{International Centre for Theoretical Physics, Strada Costiera 11, Trieste 34151, Italy}
\affiliation[4]{Physik-Department and Excellence Cluster Universe, Technische Universit\"at  M\"unchen, 85748 Garching, Germany}
\affiliation[5]{National Research Centre Kurchatov Institute, 123182 Moscow, Russia}
\affiliation[6]{Institut f\"ur Kernphysik, Forschungszentrum J\"ulich, 52425 J\"ulich, Germany}
\affiliation[7]{Dipartimento di Fisica, Universit\`a degli Studi e INFN, 20133 Milano, Italy}
\affiliation[8]{Chemical Engineering Department, Princeton University, Princeton, NJ 08544, USA}
\affiliation[9]{Institut f\"ur Experimentalphysik, Universit\"at Hamburg, 22761 Hamburg, Germany}
% ...

\emailAdd{spokesperson-borex@lngs.infn.it}

\abstract{
The Borexino detector measures solar neutrino fluxes via neutrino-electron elastic scattering. Observed spectra are determined by the solar-$\nu_{e}$ survival probability $P_{ee}(E)$, and the chiral couplings of the neutrino and electron. Some theories of physics beyond the Standard Model postulate the existence of Non-Standard Interactions (NSI's) which modify the chiral couplings and $P_{ee}(E)$. In this paper, we search for such NSI's, in particular, flavor-diagonal neutral current interactions that modify the $\nu_e e$ and $\nu_\tau e$ couplings using Borexino Phase II data.
Standard Solar Model predictions of the solar neutrino fluxes for both high- and low-metallicity assumptions are considered. 
No indication of new physics is found at the level of sensitivity of the detector and constraints on the parameters of the NSI's are placed. 
In addition, with the same dataset the value of $\sin^2\theta_W$ 
is obtained with a precision comparable to that achieved in reactor antineutrino experiments.
}

\arxivnumber{1905.03512}

\begin{document}
\maketitle
\flushbottom

\section{Introduction}
 
The study of solar neutrinos is relevant not only for probing our understanding of the Sun but also for investigating neutrino properties. 
Solar neutrino experiments, primarily SNO \cite{Ahmad:2002jz} and Super-Kamiokande  \cite{Abe:2016nxk},
together with KamLAND \cite{Eguchi:2002dm,Araki:2004mb,Abe:2008aa}, have resolved the solar neutrino problem with 
the large mixing angle (LMA) MSW flavor conversion effect \cite{Wolfenstein:1977ue,Wolfenstein:1979ni,Mikheev:1986gs,Mikheev:1986wj}. 
Improved experimental precision may reveal the effects of physics beyond the Standard Model, such as 
sterile neutrinos, particle dark matter or non-standard interactions (NSI's) of the neutrino \cite{Maltoni:2015kca,Pallavicini:2017jne,Capozzi:2017auw,Essig:2018tss}.
In this article, we present the latest sensitivity of Borexino to study the latter.

The Borexino experiment at the Laboratori Nazionali del Gran Sasso (LNGS) \cite{Alimonti:2008gc} detects solar neutrinos through the neutrino-electron elastic scattering interaction on a $\sim$280 ton liquid scintillator target with $(3.307 \pm 0.003) \times 10^{31}$ electrons per 100 ton of the mass. 
During the Phase-I period (May 16, 2007--May 8, 2010) Borexino had 740.7 live days of data taking \cite{Bellini:2013lnn,Bellini:2014uqa}.   
Following Phase-I, an extensive scintillator purification campaign was conducted resulting in significant reductions of radioactive contaminants.
Uranium-238 and Thorium-232 levels were reduced to 
${}^{238}\text{U} < 9.4 \times 10^{-20} \, \text{g/g}$ (95\% C.L.) and
${}^{232}\text{Th} < 5.7 \times 10^{-19} \, \text{g/g}$ (95\% C.L.). ${}^{85}\text{Kr}$ and ${}^{210}\text{Bi}$ concentrations were reduced by factors $\: \sim4.6$ and $\: \sim2.3$, respectively \cite{Agostini:2017ixy}. The Phase-II data, analyzed in this paper, were collected from December 14, 2011 until May 21, 2016, corresponding to 1291.51~days $\times$ 71.3~t (252.1 ton$\cdot$years) of fiducial exposure. Reduction of the background, longer exposure, and better understanding of the detector response allowed for fits to be performed in a wider energy range ($0.19\,\mathrm{MeV}<T<2.93\,\mathrm{MeV}$,
where $T$ is the recoil-electron kinetic energy) to include $pp$, ${}^{7}\text{Be}$, $pep$, and CNO electron-recoil spectra \cite{Agostini:2018uly} \footnote{The energy spectra of $pp$ and CNO neutrinos are continuous and extend up to 0.42 MeV and 1.74 MeV, respectively. ${}^{7}\text{Be}$ ($E$ = 0.384 MeV and 0.862 MeV) and $pep$ ($E$ = 1.44 MeV) neutrinos are monoenergetic.

   In Ref. \cite{Agostini:2018uly},
a high-energy region of $3.2 < T < 16$ MeV was also considered to measure ${}^{8}\text{B}$ neutrinos with a continuous energy spectrum extending up to about 16.5 MeV.}.
Taking advantage of these improvements, this paper uses the Phase-II data to investigate the parameters of non-standard interactions (NSI's) of the neutrino with increased sensitivity.

Solar neutrinos can be used to probe for physics beyond the SM
that affect neutrino interactions with the charged leptons and quarks.
In this paper, we restrict our analysis to the neutrino-flavor-diagonal NSI's that affect $\nu_e e$ and $\nu_\tau e$ interactions to which Borexino is particularly sensitive. We do not consider NSI's that affect the $\nu_\mu e$ interaction, which are strongly constrained by the $\nu_\mu e$ scattering CHARM~II experiment~\cite{Vilain:1994qy}.

Using Borexino to constrain NSI's was originally discussed by Berezhiani, Raghavan, and Rossi in Refs.~\cite{Berezhiani:1994hy, Berezhiani:2001rt}. They argued that the monochromatic nature of ${}^{7}\text{Be}$ solar neutrinos results in an electron recoil spectrum whose Compton-like shape is more sensitive to the $\nu e$ couplings than that from a continuous neutrino energy spectrum.
Following Refs.~\cite{Berezhiani:1994hy,Berezhiani:2001rt},
a purely phenomenological analysis based on Borexino Phase-I results \cite{Arpesella:2008mt}
was carried out in Ref.~\cite{Agarwalla:2012wf}, in which 
the roles of the main backgrounds were analyzed and bounds on
$\nu_e e$ and $\nu_\tau e$ NSI's obtained.

However, the analysis considered the effects of the NSI's at detection only.
High solar metallicity (HZ) was also assumed as input to the Standard Solar Model (SSM)
\cite{Bahcall:1986pf,Bahcall:1987jc,Bahcall:2000nu,Vinyoles:2016djt} to predict the ${}^{7}\text{Be}$ solar neutrino flux.

This paper updates and improves upon the analysis of Ref.~\cite{Agarwalla:2012wf} by using the Phase-II data set with the full arsenal of improved analysis tools developed by the Borexino collaboration.
NSI effects are included in both propagation and detection. 
At production the NSI's  affect
the solar-neutrino spectrum only below the Borexino threshold of $\sim$50~keV \cite{Agostini:2018uly,Vitagliano:2017odj}, and are therefore neglected.   
To account for the effect of solar metallicity, 
analyses are performed for both high- (HZ) and low-metallicity (LZ) solar models.

This paper is organized as follows.  In section~2, we review the neutrino-electron interactions in the SM and with additional effects due to NSI's, and introduce the notation.
Section~3 provides an outline of the analysis strategy and, in particular, how backgrounds and uncertainties are
handled.
Results and their discussion are presented in section~4.
A summary of the main findings is presented in section~5.

\bigskip
\noindent

\section{\boldmath $\nu e$ Elastic Scattering}

\subsection{Standard Model Interactions}

Within the SM, the elastic scattering of $\nu_\alpha$ ($\alpha=e,\mu,\tau$) on electrons proceeds via $Z$-exchange (Neutral Current, NC) and, for $\nu_e$, also via $W$-exchange (Charged Current, CC).
At momentum transfers relevant for Borexino ($Q^2\ll M_W^2,M_Z^2$), the CC and NC processes are well
approximated by the point interaction:
\begin{eqnarray}
-\mathcal{L}^{\nu e}_\mathrm{CC}
& = & \dfrac{G_F}{\sqrt{2}}
\Bigl[\,
\overline{\nu}_e\gamma_\mu\left(1-\gamma^5\right)e
\,\Bigr]\!
\Bigl[\,
\overline{e}\,\gamma^\mu\left(1-\gamma^5\right)\nu_e
\,\Bigr]
\;=\; 2\sqrt{2}\,G_F
\Bigl[\,
\overline{\nu}_{eL}\gamma_\mu\nu_{eL}
\,\Bigr]\!
\Bigl[\,
\overline{e}_L\gamma^\mu e_L
\,\Bigr]
\;,
\vphantom{\Bigg|}\cr
&&
\label{LCC}
\end{eqnarray}
\begin{eqnarray}
-\mathcal{L}^{\nu e}_\mathrm{NC} 
& = & \dfrac{G_F}{\sqrt{2}}
\Bigl[\,
\overline{\nu}_\alpha \gamma_\mu\!\left(1-\gamma^5\right)\nu_\alpha
\,\Bigr]\!
\Bigl[\,
\overline{e}\gamma^\mu\!\left(g^{\nu e}_{LV}-g^{\nu e}_{LA} \gamma^5\right)e
\,\Bigr] 
\vphantom{\Bigg|}\cr
& = & 2\sqrt{2}\,G_F
\Bigl[\,
\overline{\nu}_{\alpha L}\gamma_\mu\nu_{\alpha L}
\,\Bigr]
\Bigl[\,
 g^{\nu e}_{LL}\left(\overline{e}_L\gamma^\mu e_L\right)
+g^{\nu e}_{LR}\left(\overline{e}_R\gamma^\mu e_R\right)
\,\Bigr]
\;,
\vphantom{\Bigg|}
\label{LNC}
\end{eqnarray}
where we have used the Fierz transformation \cite{Fierz:1937,Nieves:2003in} to rewrite the CC interaction into
NC form,  and we follow the notation of the Review of Particle Physics \cite{Erler:2018} for the NC coupling constants.
The tree-level values of these couplings are
\begin{eqnarray}
g^{\nu e}_{LV} & = & -\frac{1}{2}+2\sin^2\theta_W \;,\cr
g^{\nu e}_{LA} & = & -\frac{1}{2} \;,\cr
g^{\nu e}_{LL} & = & \dfrac{1}{2}\left(g^{\nu e}_{LV} + g^{\nu e}_{LA}\right) \;=\; 
-\frac{1}{2}+\sin^2\theta_W\;,\cr
g^{\nu e}_{LR} & = & \dfrac{1}{2}\left(g^{\nu e}_{LV} - g^{\nu e}_{LA}\right) \;=\; 
\sin^2\theta_W \;.
\label{gLgR}
\end{eqnarray}
Combining \eqref{LCC} and \eqref{LNC} we have
\begin{equation}
-\mathcal{L}^{\nu_\alpha e}
\;=\;
2\sqrt{2}\,G_F
\Bigl[\,
\overline{\nu}_L\gamma_\mu\nu_L
\,\Bigr]
\Bigl[\,
 g_{\alpha L}\left(\overline{e}_L\gamma^\mu e_L\right)
+g_{\alpha R}\left(\overline{e}_R\gamma^\mu e_R\right)
\,\Bigr]
\;,
\label{nueSMinteractions}
\end{equation}
with 
\begin{eqnarray}
\label{gLgRconstants}
g_{\alpha L} & = & 
\begin{cases}
g_{LL}^{\nu e} + 1 & \mbox{for $\alpha=e$,} \\
g_{LL}^{\nu e}     & \mbox{for $\alpha=\mu,\tau$,}
\end{cases}
\cr
g_{\alpha R} & = & \;\; g_{LR}^{\nu e}\qquad\quad\mbox{for $\alpha=e,\mu,\tau$.}\vphantom{\Bigg|}
\end{eqnarray}
For a monochromatic neutrino of energy $E$ and flavor $\alpha$ 
scattering off an electron at rest, the interaction \eqref{nueSMinteractions} predicts
the spectrum of the kinetic energy $T$ of the recoiling electrons to be \cite{tHooft:1971ucy,Bahcall:1986pf}
\begin{equation}
\frac{d\sigma_{\alpha}(E,T)}{dT} \;=\; \frac{2}{\pi}\, G_{F}^2 m_e 
\left[ g_{\alpha L}^2 + 
g_{\alpha R}^2 \left(1 - \frac{T}{E}\right)^2 - g_{\alpha L} g_{\alpha R} \frac{m_e T}{E^2} 
\right]
\;,
\label{diffCS}
\end{equation}
where neutrino masses have been neglected and $T$ is constrained as:
\begin{equation}
0 \;\le\; T \;\le\; T_{\max} \;=\; \dfrac{E}{1+\dfrac{m_e}{2E}}\;.
\label{Tmax}
\end{equation}

\subsection{Radiative Corrections}

The tree-level expression for the cross section given in Eq.~\eqref{diffCS} is modified by radiative corrections
\cite{Ram:1967zza,Marciano:1980pb,Sarantakos:1982bp,Wheater:1982yk,Bahcall:1995mm,Passera:2000ug}.
In the present analysis, these corrections are accounted for following the 1995 paper of Bahcall, Kamionkowski, and Sirlin \cite{Bahcall:1995mm}
with parameters updated to reflect the more recently available experimental data, {\it e.g.} the Higgs mass,
and $\hat{s}_Z^2 = 0.23129$.\footnote{%
$\hat{s}_Z^2$ denotes the $\overline{\text{MS}}$ value of $\sin^2\theta_W$ at the $Z$-mass scale.
The value of $\hat{s}_Z^2=0.23129\pm 0.00005$ is from the 2016 Review of Particle Physics \cite{Patrignani:2016xqp}. 
It has subsequently been updated to $\hat{s}_Z^2=0.23122\pm 0.00003$ in the 2018 Review of Particle Physics \cite{Erler:2018,Tanabashi:2018oca}, but this difference is too small to be of relevance to the analysis of this paper.
}
Borexino does not distinguish between muon- and tau-neutrinos, and the difference in radiative corrections for the two flavors is consequently ignored: the radiative corrections to $\nu_{\tau}$ were set to be the same as those for the $\nu_{\mu}$.
%
The sizes of these radiative corrections are generally small compared to the experimental precision of Borexino with the exception of the recent 2.7\%  measurement of the ${}^{7}\text{Be}$ solar neutrino flux \cite{Agostini:2018uly}.
The effect of radiative corrections has a comparable magnitude, resulting in a $\sim$2\% reduction of the total cross section for $\nu_e$, and a 1.2\% increase for $\nu_{\mu / \tau}$ \cite{Bahcall:1995mm}. Nevertheless, they have little impact on the present analysis.

\subsection{Non-Standard Interactions (NSI's)}\label{NSIsection}

In addition to the SM interactions presented above, many models of physics beyond the Standard Model (BSM) predict new interactions of the neutrinos with the other SM fermions \cite{Antusch:2008tz,Gavela:2008ra,Malinsky:2008qn,Ohlsson:2009vk,Medina:2011jh,Farzan:2015doa,Farzan:2015hkd,Farzan:2016wym,Blennow:2016jkn,Deniz:2017zok,Pospelov:2017kep}.
Phenomenologically, such non-standard interactions (NSI's) of the NC type are described by the Lagrangian density \cite{Berezhiani:2001rs,Ohlsson:2012kf}

\begin{equation} 
-\mathcal{L}_{\text{NC-NSI}} \;=\; 
\sum_{\alpha,\beta}
 2 \sqrt{2} G_F\,\varepsilon_{\alpha \beta}^{f f' C} 
 \bigl(\bar{\nu}_{\alpha} \gamma^{\mu} P_L \nu_{\beta}\bigr) 
 \bigl(\bar{f} \gamma_{\mu} P_C f'\bigr)
 \;,
\label{lagNSI}
\end{equation}
where $\alpha,\beta=e,\mu,\tau$ label the neutrino flavor, 
$f$ and $f'$ are leptons or quarks of the same charge but not necessarily the same flavor, 
$C$ is the chirality of the $ff'$ current ($L$ or $R$),
and $\varepsilon_{\alpha \beta}^{f f' C}$ is a dimensionless coupling parametrizing the strength of the
NSI interaction normalized to $G_F$. 
Allowing $\alpha\neq\beta$ and $f\neq f'$ in \eqref{lagNSI} accounts for possible flavor-changing NSI's.
Hermiticity of the interaction demands 
\begin{equation}
\varepsilon_{\alpha\beta}^{ff'C} \;=\;
\varepsilon_{\beta\alpha}^{f'fC*}\;,
\end{equation}
where the asterisk denotes complex conjugation.
%
In the current analysis, however, 
we restrict our attention to the flavor-diagonal case $f=f'=e$ and $\alpha=\beta$, 
and denote $\varepsilon_{\alpha}^{C} \equiv \varepsilon_{\alpha\alpha}^{eeC}$. 
Borexino, relying on neutrino-electron elastic scattering, is particularly sensitive to this type.
A discussion on BSM models which may produce such NSI's can be found in Refs.~\cite{Berezhiani:2001rs,Davidson:2003ha,Barranco:2007ej,Farzan:2017xzy} \footnote{Before the confirmation of neutrino oscillations by the KamLAND experiment, NSI's with massless neutrinos had also been invoked to address the solar neutrino anomaly. See Refs.~\cite{Guzzo:1991hi, Guzzo:1991cp, valle1987resonant, Roulet:1991sm, Barger:1991ae,Guzzo:2001mi}.}. 

NSI's can affect neutrino production, detection, and propagation.
Inside the Sun,
the flavor diagonal NSI's under consideration
contribute to the production of same-flavor $\nu\overline{\nu}$ pairs
via 
photo-production ($\gamma e\to e\nu\overline{\nu}$), $\nu\overline{\nu}$-Bremsstrahlung
(the photon leg in $\gamma e\to e\nu\overline{\nu}$ is anchored on an ion or another electron), 
etc. \cite{Vitagliano:2017odj}.
However, the energies of the neutrinos and anti-neutrinos produced by these processes are 
expected to be in the few keV range, well below the $\sim$50~keV detection threshold
of Borexino \cite{Agostini:2018uly}.

At detection, $\varepsilon_{\alpha}^{L/R}$ ($\alpha=e,\mu,\tau$) shift the coupling constants that appear in the expression for the differential cross section, Eq.~\eqref{diffCS}:
\begin{eqnarray}
\label{gRmodif}
g_{\alpha R} & \quad\to\quad & \tilde{g}_{\alpha R} \;=\; g_{\alpha R} + \varepsilon_{\alpha}^R \;,
\\
\label{gLmodif}
g_{\alpha L} & \quad\to\quad & \tilde{g}_{\alpha L} \;=\; g_{\alpha L} + \varepsilon_{\alpha}^L \;.
\end{eqnarray}
Strong bounds on $\varepsilon_{\mu}^{L/R}$ had already been obtained by the 
$\nu_\mu e$ scattering experiment CHARM II \cite{Vilain:1994qy}, namely 
$-0.025 <  \varepsilon_{\mu}^{L} < 0.03$ and $-0.027 <  \varepsilon_{\mu}^{R} < 0.03$ at 90\% C.L. 
\cite{Davidson:2003ha}.\footnote{%
These are one-parameter-at-a-time bounds.  
One-parameter projections of two-parameter bounds at 90\% C.L. are given as 
$-0.033 <  \varepsilon_{\mu}^{L} < 0.055$ and $-0.040 <  \varepsilon_{\mu}^{R} < 0.053$ 
in Ref.~\cite{Barranco:2007ej}.}
Therefore, we restrict our attention to the remaining four parameters: $\varepsilon_{e}^{L/R}$ and $\varepsilon_{\tau}^{L/R}$. We do not consider the full 4-dimensional space $\varepsilon_{\alpha}^{L/R}$ ($\alpha = e, \tau$): such a detailed description is not necessary at the current level of sensitivity to NSI's. 
Instead, we investigate the $\varepsilon_{e}^{L/R}$ and $\varepsilon_{\tau}^{L/R}$ cases separately, even though -- as it will become evident later with \eqref{Vx} and \eqref{recoilspec} -- these groups of parameters cannot be decoupled with Borexino.\\


The description of how NSI's affect neutrino propagation can be found in Ref.~\cite{Bolanos:2008km,Farzan:2017xzy,Friedland:2004pp, Maltoni:2015kca}. 
Let us discuss in some detail what we should expect from their inclusion.
Neutrino propagation in matter is only sensitive to the vectorial combinations 
$\varepsilon_{\alpha}^{V} \equiv \varepsilon_{\alpha}^{L} + \varepsilon_{\alpha}^{R}$.
They modify the matter-effect potential in the flavor basis to
\begin{equation}
V(x)\begin{bmatrix} 1 & 0 & 0 \\ 0 & 0 & 0 \\ 0 & 0 & 0 \end{bmatrix}
\;\;\to\;\;
V(x)\begin{bmatrix} 1+\varepsilon_{e}^{V} & \ 0 & \ 0 \ \\ 0 & \ 0 & \ 0 \ \\ 0 & \ 0 & \ \varepsilon_{\tau}^{V} \end{bmatrix}
\;,
\label{MatterEffectV}
\end{equation}
where $V(x)=\sqrt{2}G_F N_e(x)$, and $N_e(x)$ is the 
electron density at location $x$.
From a practical point of view, 
$\varepsilon_{e}^{V}=\varepsilon_{e}^{L}+\varepsilon_{e}^{R}$ and 
$\varepsilon_{\tau}^{V}=\varepsilon_{\tau}^{L}+\varepsilon_{\tau}^{R}$ 
can be introduced as a shift in the matter-effect 
potential $V(x)$ in two-flavor oscillation analysis:
%
\begin{equation}
V(x) \quad\rightarrow\quad V'(x) \,=\, (1- \varepsilon') V(x)\;, 
\label{Vx}
\end{equation}
%
where $\varepsilon' = \varepsilon_{\tau}^{V}\sin^2 \theta_{23} - \varepsilon_{e}^{V}$ \cite{Bolanos:2008km}. 
The derivation of this effective potential is given in appendix~\ref{Vprime}, assuming  $|\Delta m^2_{31}| \gg 2EV(x)$, where $E$ is the neutrino energy. There, it is also shown that the first oscillation resonance occurs at 
$2EV'(x)c_{13}^2 = 2E(1-\varepsilon')V(x)c_{13}^2 \approx \Delta m_{21}^2\cos 2\theta_{12}$.


For neutrinos coming from the center of the Sun,
where the SSM predicts $N_{e}^{\odot}(r=0) \approx 10^2 N_A = 6 \times 10^{25}/\mathrm{cm^{3}}$ \cite{Bahcall:1987jc,Bahcall:2000nu}, the resonance energy is 
\begin{equation}
E_{res}^{\odot}(0) \;\approx\; \dfrac{\Delta m^2_{21}\cos 2\theta_{12}}{(1-\varepsilon')2\sqrt{2}G_F N_e^\odot(0) c_{13}^2}
\;\approx\; \dfrac{2\;\mathrm{MeV}}{(1-\varepsilon')}\;,
\label{SunMatterEffect}
\end{equation}
where for $\Delta m^2_{21}\cos 2\theta_{12}$ and $c_{13}^2 = 1-s_{13}^2$ 
we have used the central values of the global averages from Ref.~\cite{Esteban:2016qun}.

As the electron density $N_e^\odot(r)$ decreases towards the surface of the Sun, $r\to R_\odot$, the resonance energy 
$E_{res}^{\odot}(r)$ will increase.
The presence of non-zero $\varepsilon'$ will also shift the resonance energy: positive $\varepsilon'$ to higher values
and negative $\varepsilon'$ to lower values.

The MSW effect \cite{Wolfenstein:1977ue,Wolfenstein:1979ni,Mikheev:1986gs,Mikheev:1986wj} in the energy range  $E\gtrsim E_{res}^\odot(r)$ ensures a well-defined electron neutrino survival probability $P_{ee}(E)$. For lower energies, neutrino oscillates in a vacuum regime, with a smooth $P_{ee}(E)$ change in the transition region between the two regimes of oscillations. Since the energy ranges of $pp$, ${}^{7}\text{Be}$, and $pep$ neutrinos are below the resonance, the influence of matter effect for those components is small compared to that for ${}^{8}\text{B}$ neutrinos.

The mass density at the center of the Earth according to the Preliminary Reference Earth Model (PREM) \cite{PREM:1981} 
is $\rho(r=0) \approx 13\,\mathrm{g/cm^3}$, which gives us an estimate of the electron density there as 
$N_e^{\oplus}(r=0) \approx N_A\,\rho(r=0)/2  = 4\times 10^{24}/\mathrm{cm^{3}}$.
So the resonance energy of the neutrinos at the Earth's center is 
\begin{equation}
E_{res}^{\oplus}(0) \;\approx\; \dfrac{\Delta m^2_{21}\cos 2\theta_{12}}{(1-\varepsilon')2\sqrt{2}G_F N_e^\oplus(0) c_{13}^2}
\;\approx\; \dfrac{30\;\mathrm{MeV}}{(1-\varepsilon')}\;,
\label{EarthMatterEffect}
\end{equation}
and $E_{res}^{\oplus}(r)$ will grow larger as the electron density decreases toward the surface of the Earth, $r\to R_\oplus$.
From this, one can expect matter effects due to the Earth to be small for all solar neutrino components.\footnote{%
For ${}^{8}\text{B}$ neutrinos, the day-night asymmetry for LMA-MSW has been predicted to be a few percent \cite{deGouvea:1999xe,Bahcall:2001cb}, and this has been confirmed experimentally by Super-Kamiokande \cite{Fukuda:1998rq,Smy:2003jf} and SNO \cite{Ahmad:2002ka,Aharmim:2005gt}.
The sensitivity of Borexino is insufficient to detect this day-night asymmetry.}
Indeed, the day-night asymmetry at Borexino for the $\varepsilon'=0$ case 
has been predicted to be a mere fraction of a percent \cite{Aleshin:2011hu,Ioannisian:2015qwa}, and Borexino reports
$A_{dn}^{{}^{7}\text{Be}} = 0.001\pm 0.012\text{(stat)}\pm 0.007\text{(syst)}$ in Ref.~\cite{Bellini:2011yj}.
A negative value of $\varepsilon'$ could, of course, lower the resonance energy and affect this prediction
but due to the difference in the energy scales of Eqs.~\eqref{SunMatterEffect} and \eqref{EarthMatterEffect},
one expects the effect of $\varepsilon'$ would appear in the Sun first.

\begin{figure}[t]
\centering
\includegraphics[width=0.85\textwidth]{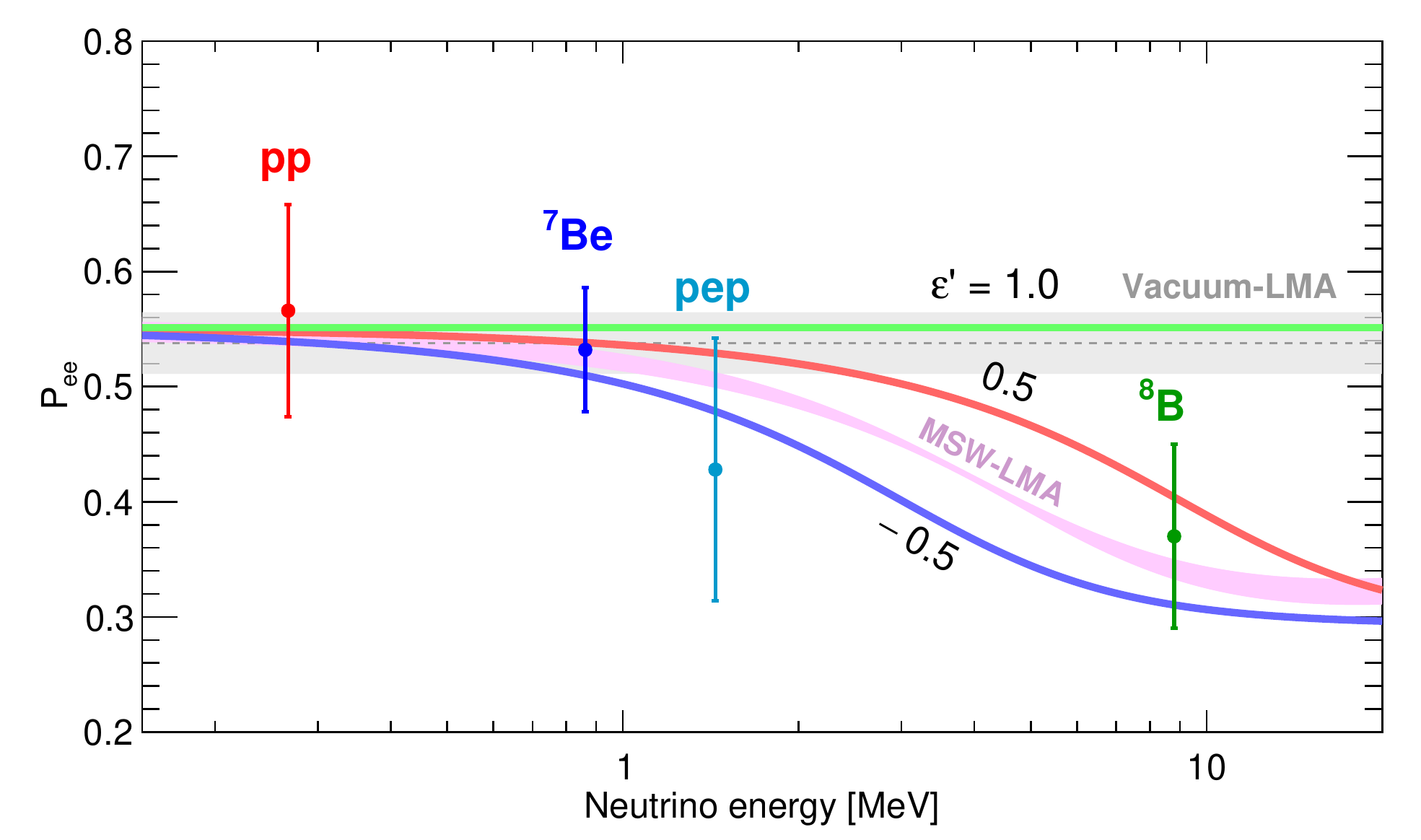}
\mycaption{
Electron neutrino survival probability $P_{ee}(E)$ as a function of neutrino energy for LMA-MSW solution with uncertainties of oscillation parameters taken into account (pink band), and LMA-MSW + NSI solutions for $\varepsilon' = - 0.5, 0.5, 1.0$ and average values of oscillation parameters. Vacuum oscillations scenario with LMA parameters is also shown (grey band). To illustrate the capability of the detector to sense NSI's at propagation, experimental points for $P_{ee}(E)$ shown for Borexino under the HZ-SSM assumption are also provided (Ref.~\cite{Agostini:2018uly}). ${}^{8}\text{B}$ and $pp$ data points are set at the mean energy of neutrinos that produce scattered electrons above the detection threshold. The error bars include experimental and theoretical uncertainties.}
\label{fig:PeeNSI}
\end{figure}

Figure~\ref{fig:PeeNSI} illustrates the effect of LMA-MSW on $P_{ee}(E)$ for several representative values of $\varepsilon'$. NSI's with $\varepsilon' > 0$  enhance $P_{ee}(E)$. For $\varepsilon' < 0$ case, $P_{ee}(E)$ is reduced.
According to Eq.~\eqref{Vx}, as $\varepsilon' \to 1$, the matter effect potential vanishes and
the $P_{ee}(E)$ tends to Vacuum-LMA scenario that assumes all solar neutrinos are oscillating in the vacuum regime. For the range between $\varepsilon' = -0.5$ and $\varepsilon' = 0.5$, the theoretically predicted shift of $P_{ee}(E)$ is within the error bars of the experimentally determined values of Borexino. The 90\% C.L. contours obtained in the present analysis are located almost entirely in this range (see  figures~\ref{fig:eLeR2d} and \ref{fig:tauLtauR2d}). Therefore, the effects of NSI's at propagation are not particularly strong, and the sensitivity to NSI's is almost entirely provided at detection.

\section{Analysis}

\subsection{Overview}



The objective of this analysis is to 
investigate the sensitivity of Borexino to the NSI parameters $\varepsilon_e^{L/R}$ and $\varepsilon_{\tau}^{L/R}$.
In contrast to the analysis of Ref.~\cite{Agostini:2018uly}, in which the $\nu e$ couplings were fixed to
those of the SM and the count rates of $pp$, ${}^7\mathrm{Be}$, and $pep$ neutrinos were fit to the data, we allow the couplings to float, assuming the SSM neutrino fluxes with either the HZ- or LZ-SSM values (table~\ref{tabfluxes}).
 

We have argued in the previous section that $\varepsilon_e^{L/R}$ and $\varepsilon_{\tau}^{L/R}$ affect neutrino propagation and detection: (i) the propagation through a shift in the matter-effect potential, Eq.~\eqref{Vx}, leading to a modification in the expected $\nu_e$ survival probability $P_{ee}(E)$,
and (ii) the detection through shifts in the effective chiral coupling constants, Eqs.~\eqref{gRmodif} and \eqref{gLmodif}, leading to modifications in the electron recoil spectra
$d\sigma_\alpha/dT$ ($\alpha=e,\tau$), Eq.~\eqref{diffCS}.


Four solar neutrino components are considered in this analysis: $pp$, ${}^{7}\text{Be}$, $pep$, and CNO.\footnote{In the present analysis, we look for deviations from the SSM + LMA-MSW predictions, so the CNO neutrino flux, 
together with the other three component fluxes, are simply fixed to those predicted by either the HZ- or LZ-SSM.} 
The SSM \cite{Bahcall:1986pf,Bahcall:1987jc,Bahcall:2000nu,Vinyoles:2016djt} predicts the energy spectra and fluxes of these neutrinos, which we denote as $d\lambda_\nu/dE$ and $\Phi_{\nu}$, where the subscript $\nu$ labels the neutrino component.

\begin{table}[t] %
\centering%
\begin{tabular}{ |c|c||c| }
\hline
Flux, $\Phi_{\nu}$ & B16(GS98)-HZ & B16(AGSS09met)-LZ  \\
\hline \hline
$pp$ & $5.98 (1 \pm 0.006)$ & $6.03(1 \pm 0.005)$   \\
$pep$ & $1.44 (1 \pm 0.01)$ & $1.46 (1 \pm 0.009)$  \\
${}^{7}\text{Be}$ & $4.93 (1 \pm 0.06)$ & $4.50 (1 \pm 0.06)$ \\
CNO & $4.88 (1 \pm 0.11)$ & $3.51 (1 \pm 0.10)$  \\
\hline 
\end {tabular}
\mycaption{The fluxes predicted by HZ- and LZ-SSM's (Ref. \cite{Vinyoles:2016djt}) and used in this analysis. Units are: $10^{10}$~($pp$), $10^9$~(${}^{7}\text{Be}$), $10^8$~($pep$, CNO) $\text{cm}^{-2} \text{s}^{-1}$.} 
\label{tabfluxes}
\end{table}

The monochromatic ${}^{7}\text{Be}$-component plays a fundamental role in this analysis. Both the shape and the normalization of the electron-recoil spectrum is well-constrained in the fit. Together with the 6\%-uncertainty in the theoretical ${}^{7}\text{Be}$ neutrino flux, it provides the highest sensitivity to NSI's among all the neutrino components.
  We do not use ${}^{8}\text{B}$ neutrinos to place bounds on NSI's. The rate of ${}^{8}\text{B}$ neutrino events cannot be determined with the spectral fit used in this analysis, being small and hidden by backgrounds in the energy region considered. Moreover, the relatively large 12\%-uncertainty on the ${}^{8}\text{B}$ neutrino flux predicted by the SSM limits its utility for this work.   
  
 Taking into account the oscillations of $\nu_{e}$ into $\nu_{\mu}$ and $\nu_{\tau}$,
the recoil spectrum for each solar neutrino component is given by
\begin{equation}
\label{recoilspec}
\frac{dR_{\nu}}{dT} 
\;=\; N_{e} \Phi_{\nu} \int dE\, \dfrac{d \lambda_{\nu}}{dE} 
\left[ 
\dfrac{d\sigma_{e}}{dT} P_{ee}(E) + 
\left(c^2_{23} \frac{d\sigma_{\mu}}{dT}+ s^2_{23} \frac{d \sigma_{\tau}}{dT}\right)
\left(1 - P_{ee}(E)\right)
\right]
\;.
\end{equation}
Here, $N_{e}$ is the number of electrons in the fiducial volume of the detector, 
$s_{23}^2 \equiv \sin^2 \theta_{23} $, and $c_{23}^2 \equiv \cos^2 \theta_{23}$. 
$\Phi_{\nu}$ is the expected total flux  
of solar neutrino component $\nu$ at the Earth, and $d\lambda_{\nu}/dE$ is the corresponding differential neutrino energy spectrum.
$P_{ee}(E)$ is the solar-$\nu_e$ survival probability to which NSI effects at propagation
have been added.
The effect of the NSI's at detection is included in the differential cross sections $d\sigma_e/dT$ and
$d\sigma_\tau/dT$, with the $\varepsilon_{e}^{L/R}$ and $\varepsilon_{\tau}^{L/R}$ parameters always combined in the recoil spectrum of Eq.~\eqref{recoilspec}.

\begin{figure}
\centering
\includegraphics[width=0.85\textwidth]{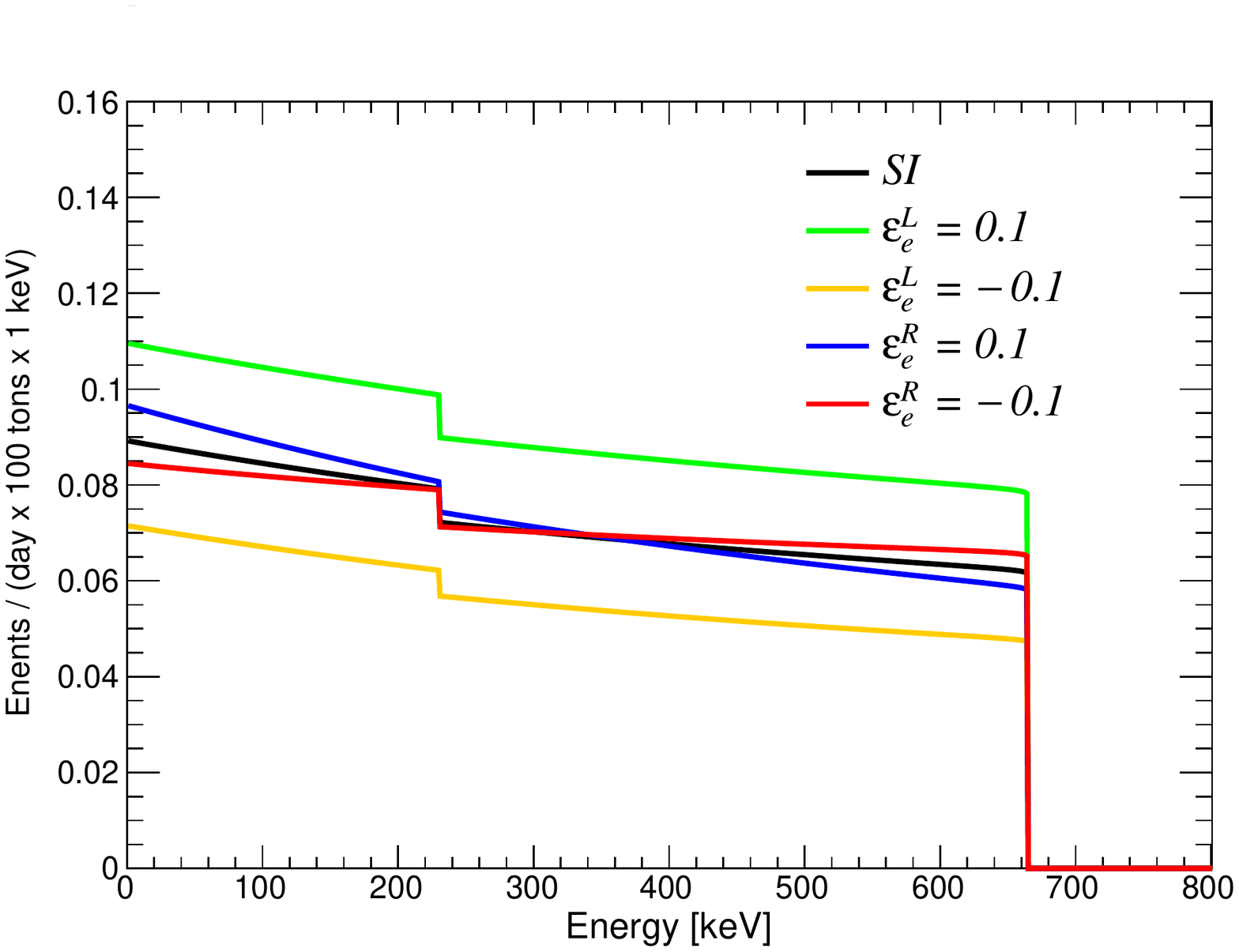}
\mycaption{The distortion of the electron recoil spectrum, Eq.~\eqref{recoilspec}, 
for the two monochromatic ${}^{7}\text{Be}$ solar neutrino lines ($E=0.384\,\mathrm{MeV}$ 
and $0.862\,\mathrm{MeV}$) due to non-zero values of $\varepsilon_{e}^{L}$ 
and $\varepsilon_{e}^{R}$. The effect of the finite energy resolution of the detector 
is not included.
}
\label{fig:modifspectra}
\end{figure} 

\begin{figure}
\centering
\includegraphics[width=0.49\textwidth]{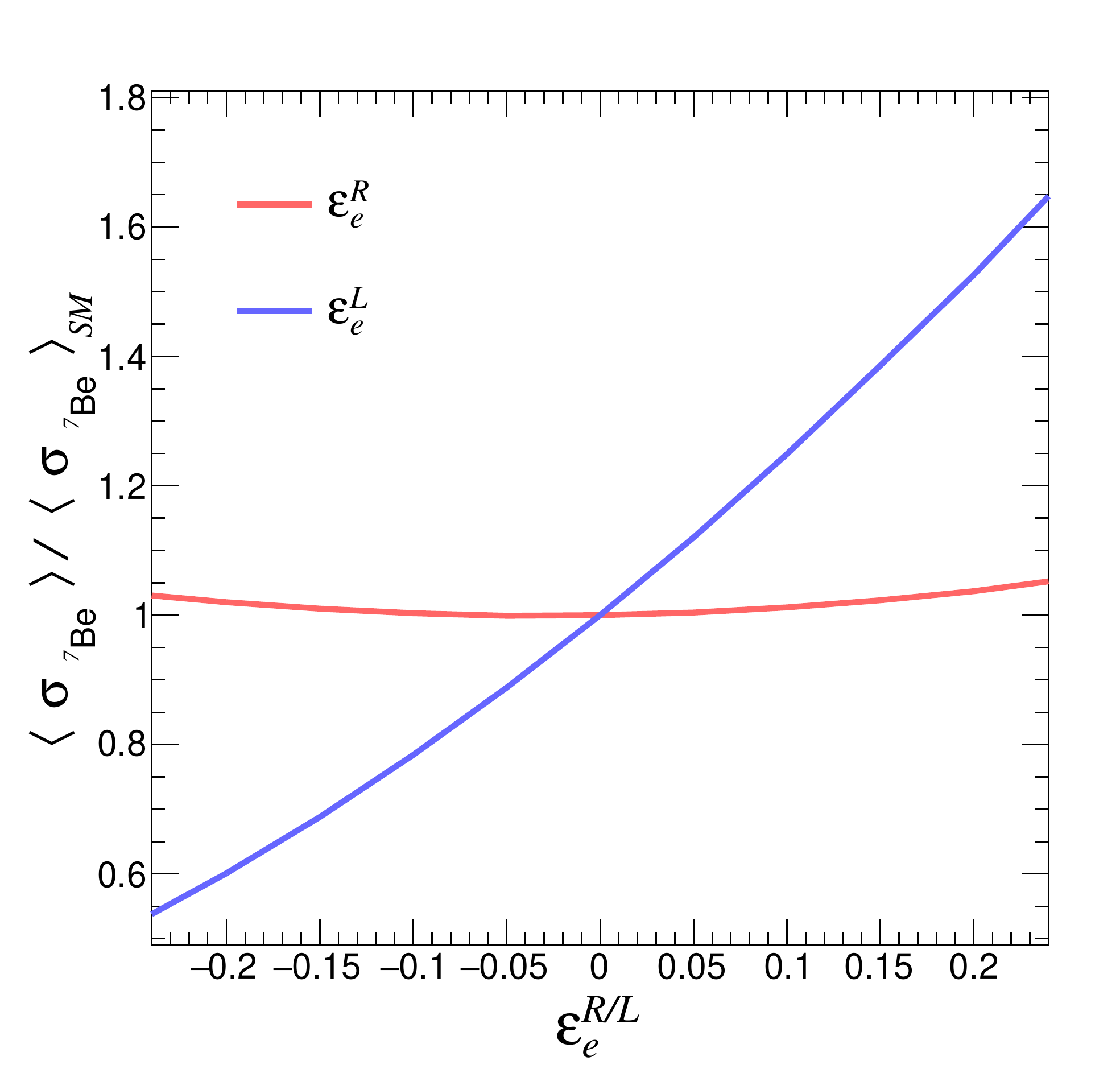}
\includegraphics[width=0.49\textwidth]{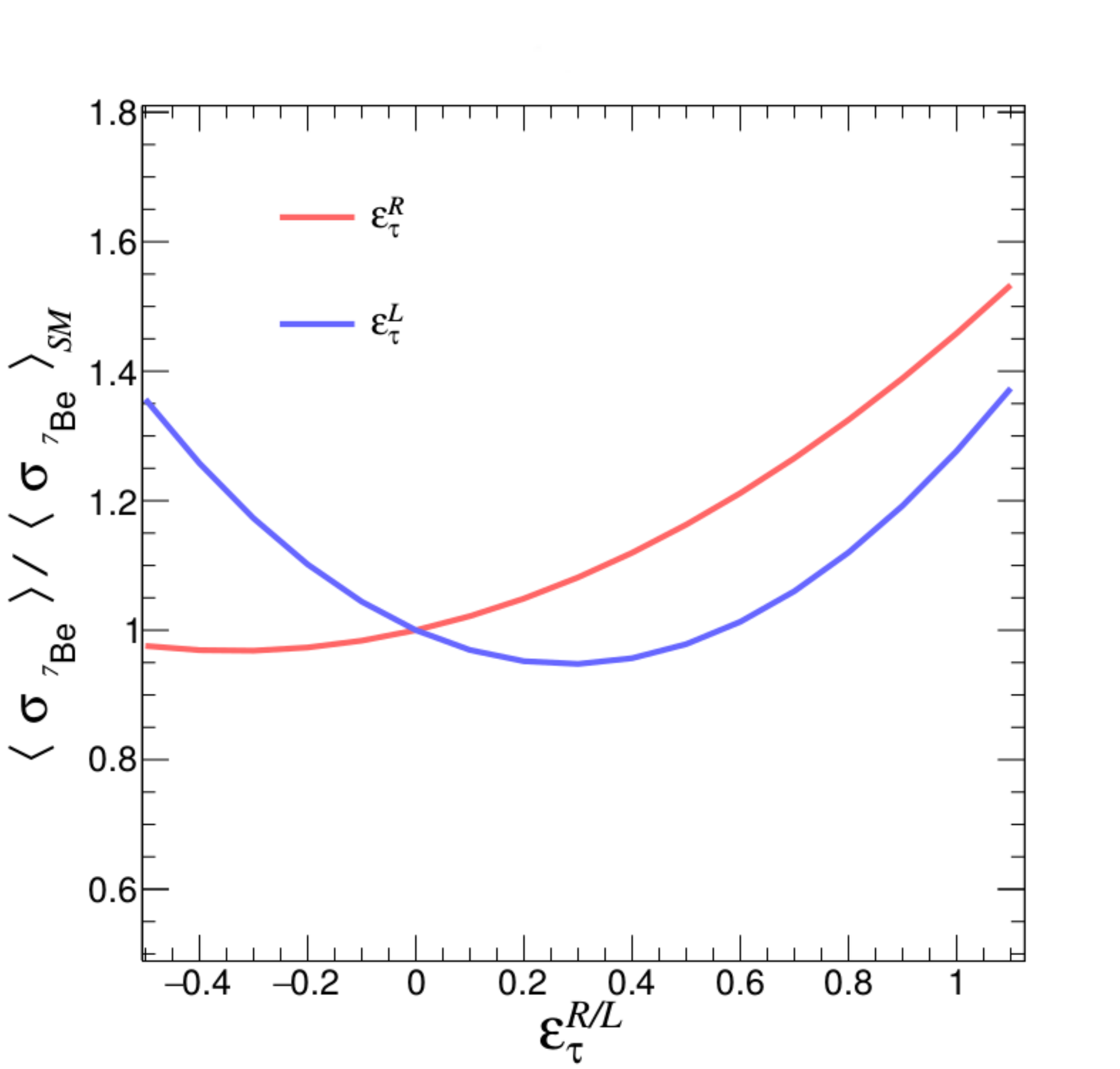}
\mycaption{The relative change of the total cross section ratio 
$\langle\sigma_{{}^{7}\text{Be}} \rangle /\langle\sigma_{{}^{7}\text{Be}} \rangle_{\mathrm{SM}}$ as function of 
$\varepsilon_{e}^{R/L}$ (left panel) and $\varepsilon_{\tau}^{R/L}$ (right panel). 
}
\label{fig:TotalCSeReL}
\end{figure}

The dependence of the ${}^{7}\text{Be}$ electron recoil spectrum 
$dR_{\mathrm{Be7}}/dT$ on the NSI's for several values of 
$\varepsilon_{e}^{R}$ and $\varepsilon_{e}^{L}$ is illustrated in figure~\ref{fig:modifspectra}. 
Note that $\varepsilon_{e}^{L}$ mostly modifies the normalization of the spectrum, while $\varepsilon_{e}^{R}$ modifies its slope. 
$\varepsilon_{\tau}^{L}$ and $\varepsilon_{\tau}^{R}$ require much larger magnitudes to achieve the same effects due to the smaller contribution of $\nu_\tau$ to $dR_{\mathrm{Be7}}/dT$. 

Integrating Eq.~\eqref{recoilspec}, one obtains a relation between the total experimental event rate $R_{\nu}$, the solar neutrino flux $\Phi_{\nu}$, and the total cross section $\langle \sigma_{\nu} \rangle$:
\begin{equation}
\label{rate}
R_{\nu} \;=\; \int \frac{d R_{\nu}}{dT} \, dT \;=\; N_{e} \Phi_{\nu}  \langle \sigma_{\nu} \rangle \;.
\end{equation} 
NSI effects at propagation and detection are both included in
the total cross section $\langle \sigma_{\nu} \rangle$.
Denoting the total cross section in the absence of NSI's as $\langle\sigma_{\nu} \rangle_{\mathrm{SM}}$,
we plot the change in the ratio $\langle \sigma_{\nu} \rangle / \langle \sigma_{\nu} \rangle_{\mathrm{SM}}$ 
for the ${}^7\mathrm{Be}$ neutrinos
due to the presence of $\varepsilon_e^L$ and $\varepsilon_e^R$ in figure~\ref{fig:TotalCSeReL}. 
Again, we see that $\varepsilon_e^L$ affects the normalization of the cross section, while
$\varepsilon_e^R$ does not.
Thus $\varepsilon_e^L$ is mostly constrained by the normalization of the cross section,
while $\varepsilon_e^R$ is mostly constrained by the shape of the recoil spectrum.

\subsection{Detector Model and Choice of Parameters}


We performed the selection of the events according to Ref.~\cite{Agostini:2018uly}, using a spherical fiducial volume to which the top and bottom polar regions are cut off: 
$R < 2.8 \, \text{m}$, and $-1.8\,\text{m}<z<2.2\,\text{m}$. 
To model the detector response, we use the analytical model of the Borexino detector discussed in detail in Ref.~\cite{Agostini:2017ixy}.
The model uses the number of triggered PMT's,  $N_p^{\,dt_{1}}$, 
within a fixed time interval $dt_1 = 230\,\mathrm{ns}$ as the estimator 
of the electron recoil energy $T$.
Various model parameters have been fixed utilizing independent measurements, 
or tuned using the Borexino Monte Carlo \cite{agostini2018monte} and calibrations \cite{back2012borexino}, while some have been left free to float in the fit.
The floating parameters include (i) the light yield, which determines the energy scale, 
(ii) two parameters for energy resolution, 
(iii) two parameters for the position and the width of the ${}^{210}\text{Po}$-$\alpha$ peak, and 
(iv) one parameter for the starting point of the ${}^{11}\text{C}$ $\beta^{+}$-spectrum.
The detector response function convoluted with the cross section $dR_\nu/dT$ provides the
functional form to be fit to the data. 



Throughout the minimization procedure, the neutrino oscillation parameters are fixed to the central values of the global fit to all oscillation data given in Ref.~\cite{Esteban:2016qun}\footnote{Strictly speaking, to use the Borexino data to constrain possible new physics effects we should not be comparing the data to the global average of Ref.~\cite{Esteban:2016qun}, which includes both Borexino Phase-I and Phase-II data in its fit.However, the numerical difference from the global average of Ref.~\cite{Capozzi:2016rtj}, which includes neither Borexino Phase I nor Phase II data, is small and does not affect the present analysis.}. Their uncertainties are directly propagated as the uncertainties of predicted neutrino rates. More details on how the uncertainties are treated can be found in section \ref{fittingprocedure}.

 For the $\varepsilon_e^{L/R}$ analysis, we only need



\begin{equation}
\label{deltaM}
\Delta m^2_{21} \,=\, m_2^2 - m_1^2 \,=\, 7.50^{+0.19}_{-0.17} \times 10^{-5}\,
\end{equation} 

\begin{equation} 
\label{sin12}
\sin^2\theta_{12} \,=\, 0.306^{+0.012}_{-0.012}\;,
\end{equation} 

\begin{equation} 
\label{sin13}
\sin^2\theta_{13} \,=\, 0.02166^{+0.00077}_{-0.00077}\;,
\end{equation} 
which are valid for any choice of neutrino mass hierarchy.

It is worthwhile to mention that the measurements of $\theta_{12}$ and $\Delta m_{21}^2$ from the global oscillation data may be altered if we consider NSI's in the fit. The solar neutrino experiments such as Super-Kamiokande and SNO provide crucial information on $\theta_{12}$ and $\Delta m_{21}^2$, and at the same time they are sensitive to the same NSI's in propagation and detection that we consider in this paper. However, the reactor experiment KamLAND is unaffected by flavor-diagonal neutral current NSI's involving neutrinos and electrons. Therefore, when we consider the KamLAND data along with the solar neutrino data mostly coming from Super-Kamiokande and SNO, the solar oscillation parameters $\theta_{12}$, $\Delta m_{21}^2$ remain robust even in the presence the NSI's discussed in this paper. In fact, it was shown in Ref. \cite{Bolanos:2008km} that Super-Kamiokande and SNO can place competitive constraints on $\varepsilon_{e}^{L}$ and $\varepsilon_{e}^{R}$ with the help of KamLAND data which provides NSI-independent measurement of $\theta_{12}$ and $\Delta m_{21}^2$. Note also that for the $\varepsilon_{e}^{L/R}$ analysis $d\sigma_\mu/dT = d\sigma_\tau/dT$ 
when $\varepsilon_{\tau}^{L/R}=0$, and Borexino is insensitive to the value of $\theta_{23}$.

For the $\varepsilon_{\tau}^{L/R}$ analysis, we also need to specify $\theta_{23}$.
The 1$\sigma$ ranges given in Ref.~\cite{Esteban:2016qun} for Normal and Inverted Hierarchies are
%
\begin{equation}
\sin^2\theta_{23}
\;=\;
\begin{cases}
\;0.441^{+0.027}_{-0.021} & \quad \mathrm{NH} \\
\;0.587^{+0.020}_{-0.024} & \quad \mathrm{IH}
\end{cases}
\end{equation}
%
It is easy to see that $\sin^2\theta_{23}$ is included linearly in expression \eqref{recoilspec}, and the sensitivity to $\varepsilon_{\tau}^{L/R}$ is proportional to its value. To obtain a conservative limit, we fix $\sin^2\theta_{23}$ to the NH value and propagate its uncertainty into systematic error together with other oscillation parameters. 






\begin{figure}
\centering
\includegraphics[width=1.0\textwidth]{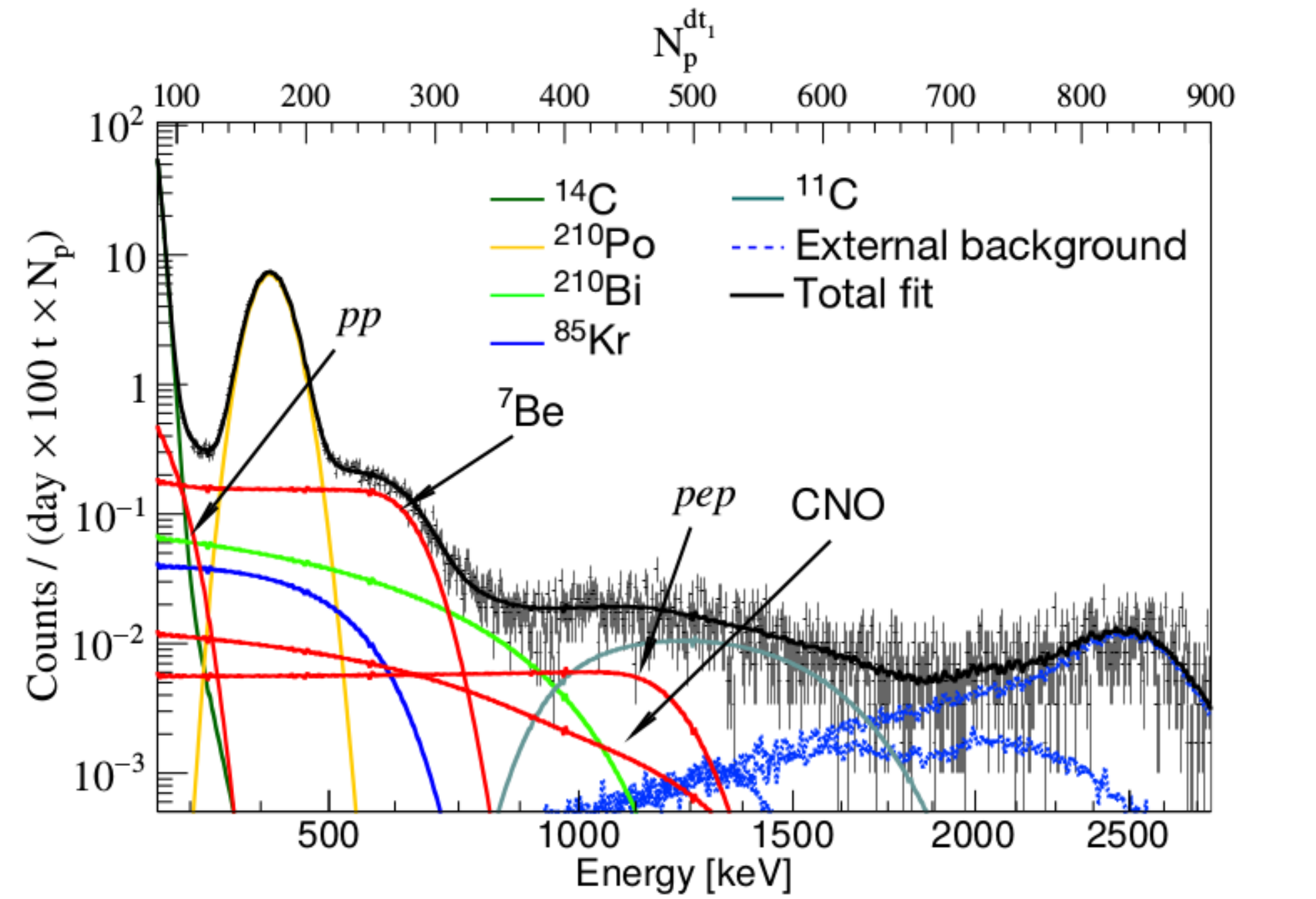}
\mycaption{Example of fit of the Borexino energy spectrum. The fit was performed using the $N_p^{dt_1}$ energy estimator. The bottom horizontal axis has been converted from $N_p^{dt_1}$ into units of energy; $N_p$ is a number of photoelectrons in the acquisition time window.}
\label{fig:fitexamples}
\end{figure}

\subsection{Backgrounds}


Radioactive contaminants lead to backgrounds that must be clearly understood
to extract unambiguous conclusions from the Borexino data.
The most recent fit of signal+background
to the observed electron recoil spectrum can be found in Ref.~\cite{Agostini:2018uly},
where the SM couplings were assumed and the event rates of three solar neutrino components 
($pp$, ${}^7\mathrm{Be}$, and $pep$) were allowed to float. 
An example fit to the experimental spectrum is shown in figure~\ref{fig:fitexamples}.
A full description of the Borexino spectral components and backgrounds is found in Ref.~\cite{Bellini:2013lnn}. 
Here, we focus on the components which are the most relevant for the current analysis:

\begin{itemize}
\item At low-energies the $\beta$-emitter ${}^{14}\text{C}$ with $Q = 156 \: \text{keV}$ is 
the main background for $pp$ neutrinos ($T_{\max} =  261 \: \text{keV}$). 

The ${}^{14}\text{C}$ contribution is constrained in the fit with an 
independent measurement by selection of events with low energy threshold. Since the rate of ${}^{14}\text{C}$ is high compared to the other components, pile-up events need to be taken into account. The detailed data selection and analysis procedures are found in Ref.~\cite{Bellini:2014uqa}. 

\item
Decays of ${}^{85}\text{Kr}$ ($\beta^{-}$,  $Q = 687 \: \text{keV}$), ${}^{210}\text{Bi}$ ($\beta^{-}$, $Q = 1160  \: \text{keV}$), and  ${}^{210} \text{Po}$ ($\alpha$, $E = 5.3 \: \text{MeV}$) are the main backgrounds for the detection of the electron recoil spectra from the two mono-energetic ${}^{7}\text{Be}$ solar neutrino lines ($E = 384 \: \text{keV}$ and $862 \: \text{keV}$).

The ${}^{210}\text{Po}$ $\alpha$-decay peak ($E = 5.3$ MeV) appears at $\sim$400 keV due to ionization quenching effects in the liquid scintillator. While very intense with respect to the other spectral components, the shape of the polonium peak is very distinct, well understood, and easily separable in the fit. 

The $\beta$ spectra of ${}^{210}\text{Bi}$ and ${}^{85}\text{Kr}$ overlap with the
${}^{7}\text{Be}$ electron-recoil spectrum leading to a modification of its shape. 
This reduces the sensitivity to the right-handed NSI parameter $\varepsilon_{\alpha}^{R}$. 
The background from ${}^{85} \text{Kr}$ is quite serious since the shape of its $\beta$-spectrum and its end-point are close to the step-like spectrum of ${}^{7}\text{Be}$. 

%


\item Other backgrounds necessary to the fit of the experimental spectrum are cosmogenic $\beta^+$ emitter ${}^{11}\text{C}$, and $\gamma$-rays from ${}^{208}\text{Tl}$, ${}^{214}\text{Bi}$, and ${}^{40}\text{K}$ from components of the detector external to the scintillator. 

\end{itemize}


\subsection{Fit Procedure}
\label{fittingprocedure}

The fitting procedure consists of the multivariate maximization of the composite likelihood function $\mathcal{L}(\vec{k}\,|\varepsilon,\vec{\theta})$, specifically developed to be able to detect $pep$, and CNO neutrinos hidden by the cosmogenic $\beta^+$ ${}^{11}\text{C}$ and external  backgrounds:
\begin{equation}
\mathcal{L}(\vec{k}\,|\varepsilon,\vec{\theta}) \;=\; \mathcal{L}^{TFC}_{sub}(\vec{k}\,|\varepsilon,\vec{\theta}) \cdot \mathcal{L}^{TFC}_{tag}(\vec{k}\,|\varepsilon,\vec{\theta}) \cdot \mathcal{L}_{P}( \vec{k}\,|\vec{\theta}) \cdot \mathcal{L}_{R}(\vec{k}\,|\vec{\theta}).
\label{LikelihoodFunction}
\end{equation}
Here, $\varepsilon$ is the NSI parameter we would like to constrain, and the vector $\vec{\theta}$ collectively represents all the other model parameters of the fit, including the rates of the four solar neutrino components $R_{\nu}$, the intensities of the backgrounds, detector response parameters, etc., and $ \vec{k}$ denotes the set of experimental data.

In order to deal with ${}^{11}\text{C}$ background, the dataset was divided into two parts by the so-called three-fold coincidence (TFC) technique (Refs.~\cite{Bellini:2013lnn,Agostini:2017ixy}). The method tags events correlated in space and time with a parent muon and one or several neutrons often produced together with ${}^{11}\text{C}$. The division is based on the probability for an event to be ${}^{11}\text{C}$ and results in ${}^{11}\text{C}$-depleted (TFC-subtracted) and ${}^{11}\text{C}$-enriched (TFC-tagged) data samples. The first and the second factors of Eq.~\eqref{LikelihoodFunction} represent two separate likelihoods for TFC-subtracted and TFC-tagged experimental spectra, respectively. 
They are a standard Poisson likelihood:
\begin{equation}
\mathcal{L}^{TFC}_{sub, \: tag}(\vec{k}\,|\varepsilon,\vec{\theta}) = \prod_{i=1}^{N_E} \frac{\lambda_i (\varepsilon,\vec{\theta})^{k_i} e^{-\lambda_i(\varepsilon,\vec{\theta})}}{k_i!}
\end{equation}
where $N_E$ is the number of energy bins, $\lambda_i (\varepsilon,\vec{\theta})$ is the expected number of events in the $i$-th bin for a given set of parameters $\varepsilon$ and $\vec{\theta}$, and $k_i$ is the measured number of events in the $i$-th bin.

The residual events from ${}^{11}\text{C}$ in the TFC-subtracted spectrum can be discriminated by the algorithm incorporated into $\mathcal{L}_{P}(\vec{k}\,|\vec{\theta})$. 
To account for external backgrounds which penetrate into the fiducial volume, the fit of the spatial radial distribution of events is incorporated by $\mathcal{L}_{R}(\vec{k}\,|\vec{\theta})$.  
The more detailed description of the likelihood function and the fitting procedure can be found in section~XXI of Ref.~\cite{Bellini:2013lnn}, and in Ref. \cite{Agostini:2017ixy}.

We add penalty factors to $\mathcal{L}(\vec{k}\,|\varepsilon,\vec{\theta})$ to constrain the four neutrino rates to the SSM prediction \cite{Bahcall:1986pf,Bahcall:1987jc,Bahcall:2000nu,Vinyoles:2016djt}:
\begin{equation}
\label{penaltylikelihood}
\mathcal{L} (\vec{k}\,|\varepsilon,\vec{\theta}) \quad\rightarrow\quad \mathcal{L} (\vec{k}\,|\varepsilon,\vec{\theta}) \cdot \prod_{\nu} \exp{ \left[ -\frac{  \Big( {\theta}_{\nu} - R_{\nu}^{\text{SSM}}(\varepsilon) \Big)^2  }{ 2 \, \Big( \delta_{ R_{\nu}^{\text{SSM}} }(\varepsilon) \Big) ^2 }\right]}
\;,
\end{equation}
where $\theta_\nu$ represents the floating value of $R_\nu$. $R_\nu^{\mathrm{SSM}}(\varepsilon)$ is the expected rate calculated by \eqref{rate} from the prediction of the SSM with either the HZ or LZ assumption and for a given set of NSI parameters $\varepsilon$, and oscillation parameters fixed to their central values. $\delta_{R_\nu^{\mathrm{SSM}}}(\varepsilon)$ is its uncertainty stemming from theoretical uncertainties of the SSM and systematic uncertainties on the estimated number of target electrons $N_{e}$, on the fiducial volume, and on the oscillation parameters.

%
%
%

Performing a series of fits for different values of $\varepsilon$, one can obtain a likelihood probability distribution 
\begin{equation}
p(\varepsilon) \;=\; 
\dfrac{ \mathcal{L}(\vec{k}\,|\varepsilon,\vec{\theta}_{\max}(\varepsilon))}{
\int d \bar{\varepsilon} \, \mathcal{L} (\vec{k}\,|\bar{\varepsilon},\vec{\theta}_{\max}(\bar{\varepsilon}))}\;, 
\end{equation}
where $\vec{\theta}_{\max}(\varepsilon)$ is the set of values of $\vec{\theta}$ that
maximizes the likelihood for a particular value of $\varepsilon$.
The upper $\varepsilon_{\mathrm{up}}$ and lower $\varepsilon_{\mathrm{low}}$ bounds for a given confidence level (C.L.) can be numerically obtained by integrating the tails of the following distribution:
\begin{equation}
\int_{-\infty}^{\varepsilon_{\mathrm{low}}} d\varepsilon \: p(\varepsilon) 
\;=\; \int_{\varepsilon_{\mathrm{up}}}^{\infty} d\varepsilon \: p(\varepsilon) 
\;=\; \frac{\mathrm{1-C.L.}}{2} 
\;.
\label{statmethod1d}
\end{equation}
For the two dimensional case when two parameters ($\varepsilon_1,\varepsilon_2$) are under investigation, the confidence region is formed by the isocontour $p_{0}=\mathrm{const}$, defined though the integral over the excluded region:
\begin{equation}
\iint \limits_{p(\varepsilon_1,\varepsilon_2)<p_{0}} d\varepsilon_1 d\varepsilon_2 \: p(\varepsilon_1, \varepsilon_2) 
\;=\;  1-\mathrm{C.L.} 
\;,
\end{equation}
where $p(\varepsilon_1,\varepsilon_2)<p_{0}$ stands for the region outside of the isocontour $p_{0}$.

\section{Results}

\begin{figure}[t]
\centering
\includegraphics[width=0.49\textwidth]{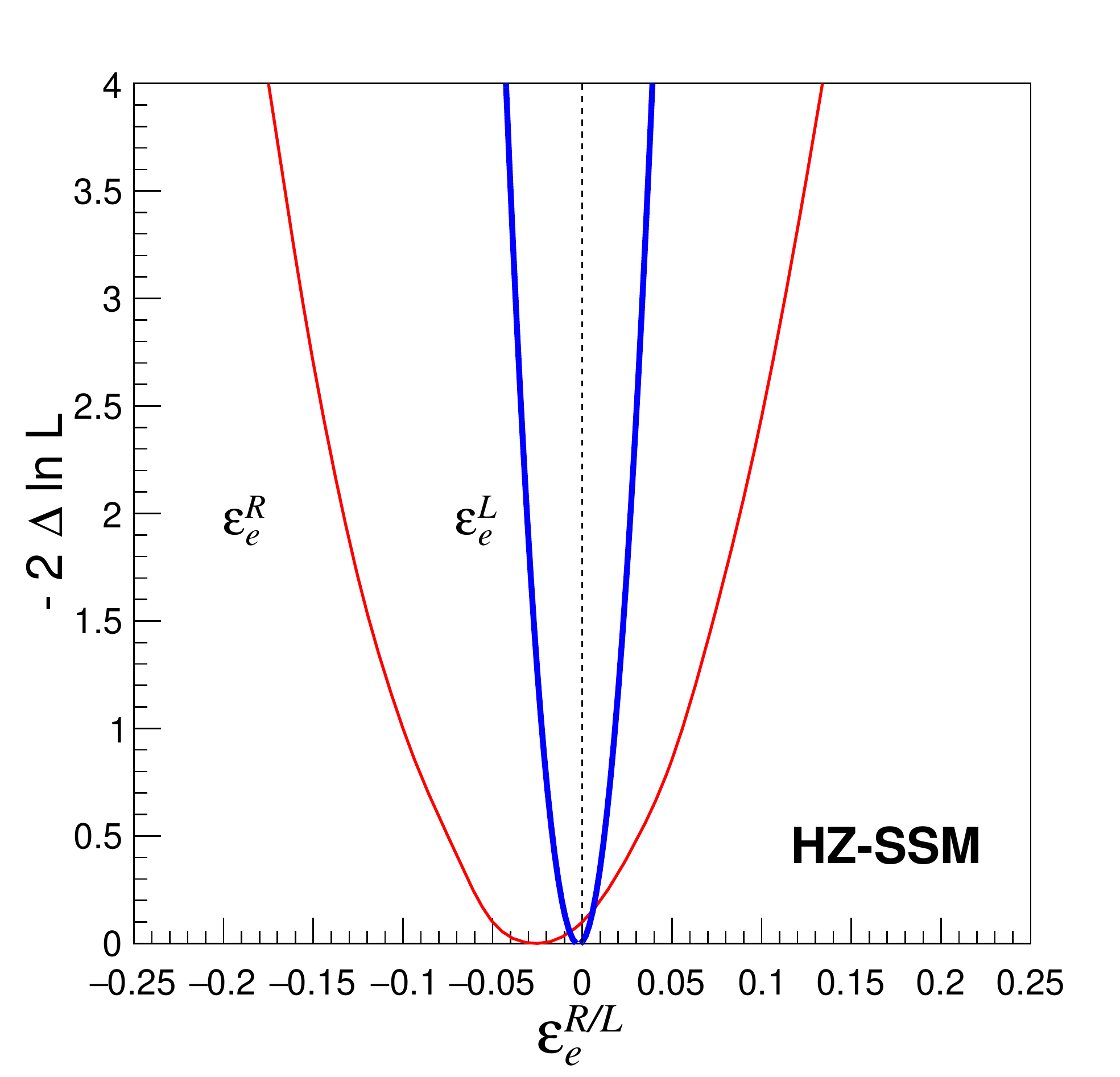}
\includegraphics[width=0.49\textwidth]{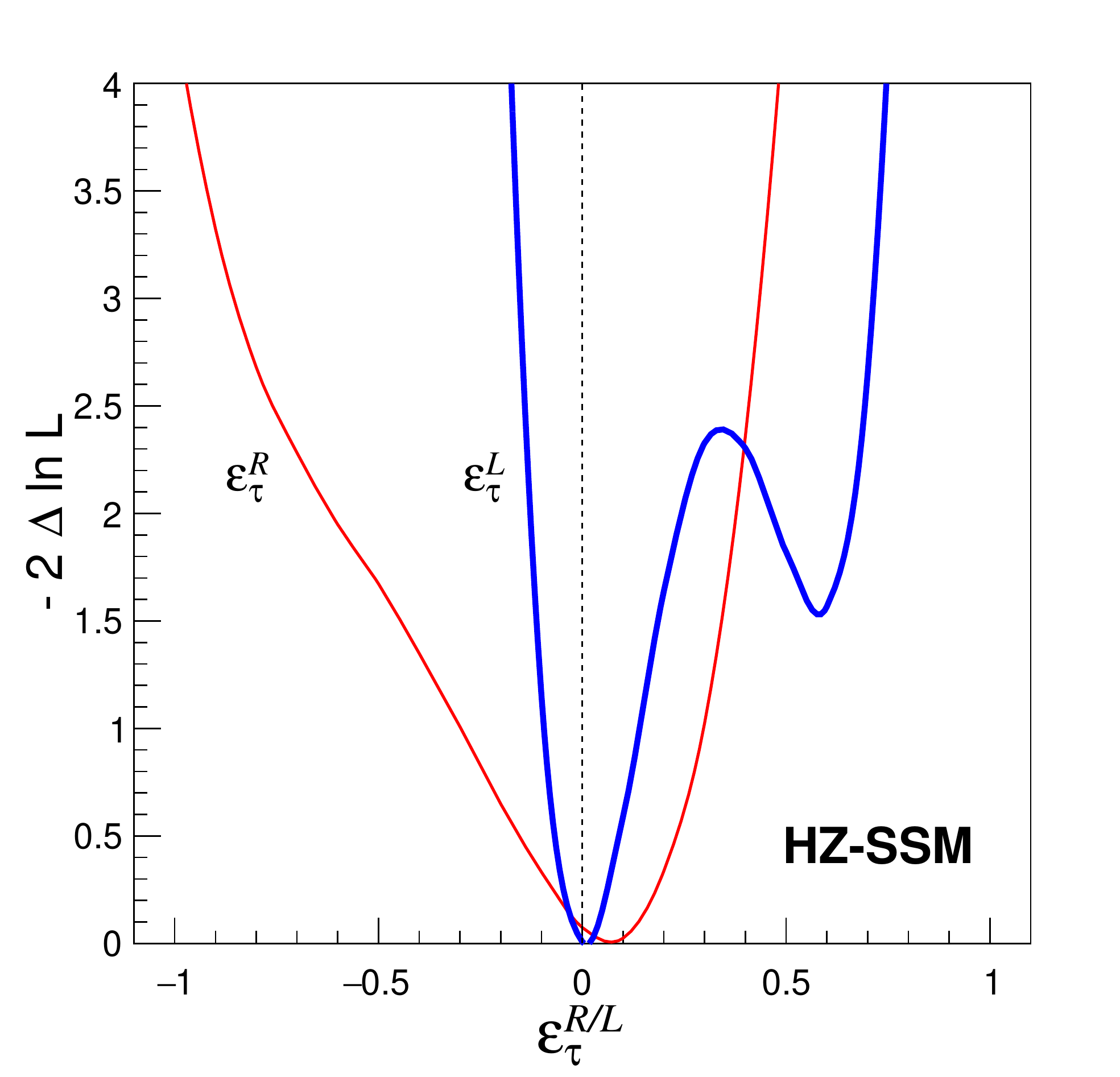}
\includegraphics[width=0.49\textwidth]{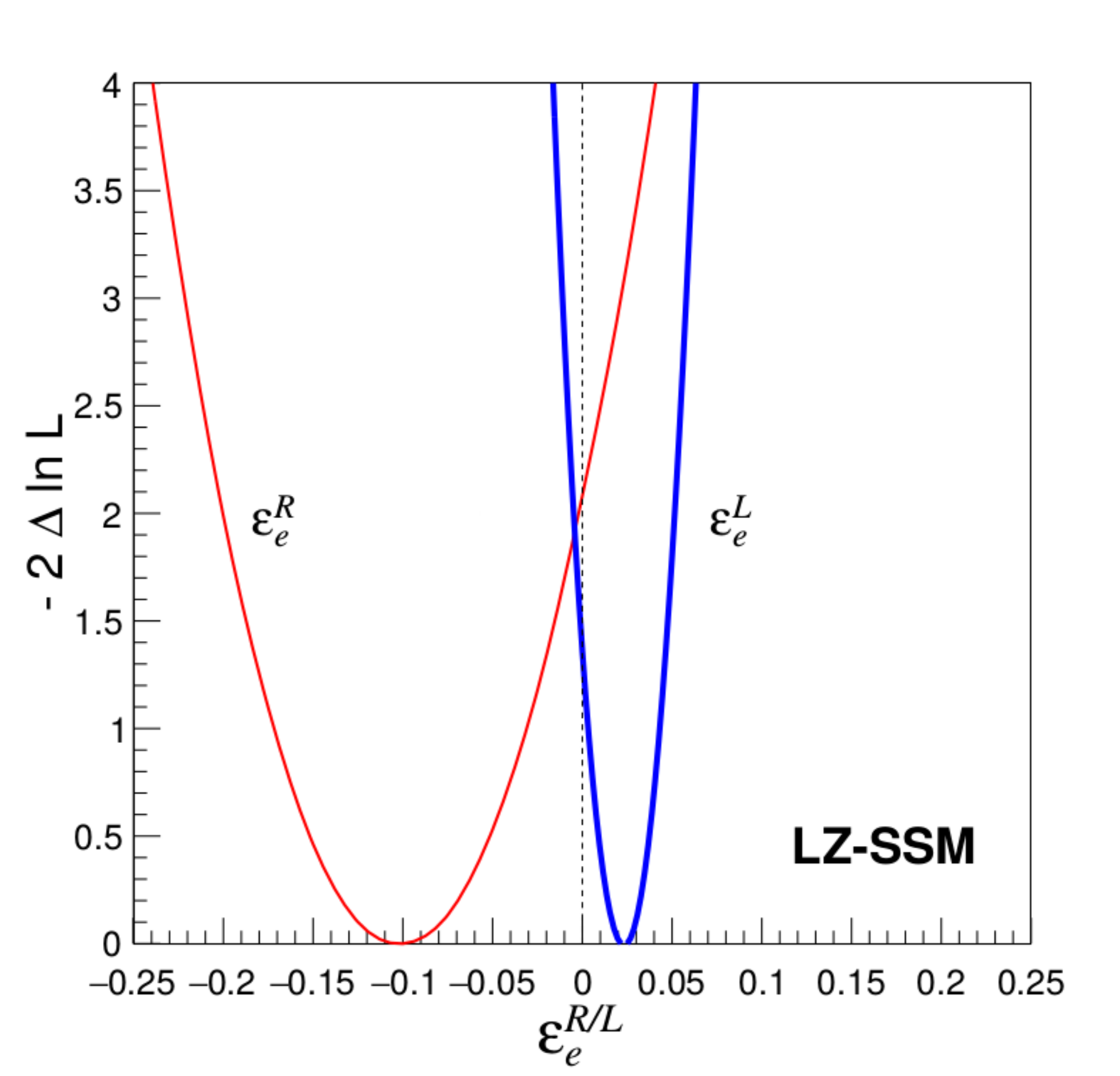}
\includegraphics[width=0.49\textwidth]{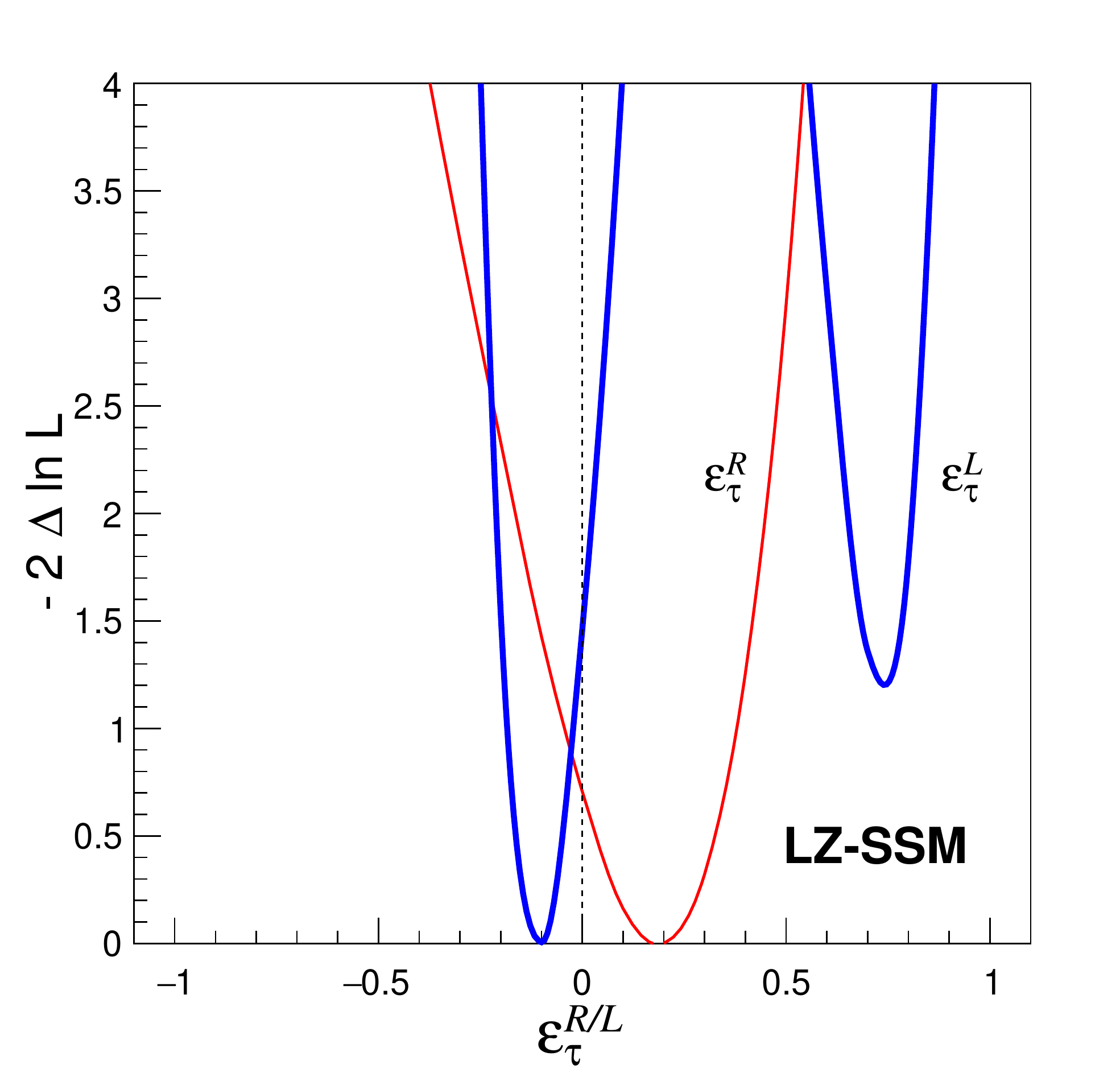}
\mycaption{Left panels show the log-likelihood profiles for 
the NSI parameter $\varepsilon_{e}^{R}$ (red line) and 
$\varepsilon_{e}^{L}$ (blue line) assuming HZ (top panel) and LZ (bottom panel) SSM's. Right panels depict the 
same for $\varepsilon_{\tau}^{R}$ (red line) and 
$\varepsilon_{\tau}^{L}$ (blue line). The profiles were obtained considering one NSI parameter at-a-time, while remaining NSI parameters were fixed to zero. 
}
\label{fig:1d-profiles}
\end{figure}

\subsection{Bounds on NSI Parameters}

In this section, we present our results.
Left panels of figure~\ref{fig:1d-profiles} shows the one-dimensional 
log-likelihood profiles for $\varepsilon_{e}^{R}$ (red curve) and
$\varepsilon_{e}^{L}$ (blue curve) assuming HZ- (top panel) and LZ-SSM (bottom panel). Right panels of figure~\ref{fig:1d-profiles}
portrays the same for $\varepsilon_{\tau}^{R}$ (red curve) and
$\varepsilon_{\tau}^{L}$ (blue curve).

Let us first discuss the HZ-SSM case (top panels). One can see that the sensitivity of Borexino to the NSI parameter $\varepsilon_{e}^{L}$ is more pronounced as compared to its sensitivity to $\varepsilon_{e}^{R}$ (see top left panel of figure~\ref{fig:1d-profiles}). 
The main reason behind this is that the normalization of neutrino 
events is well determined by the fit, which in turn provides competitive 
constraints for $\varepsilon_{e}^{L}$. In contrast, the fit still permits 
quite a wide range for $\varepsilon_{e}^{R}$, since the possible modification 
in the shape of the event spectra due to non-zero $\varepsilon_{e}^{R}$
can be easily mimicked by the principle background components (mainly ${}^{85}\text{Kr}$) discussed above. Note that the minima of the
one-dimensional log-likelihood profiles for $\varepsilon_{e}^{R}$ (red line in left panel) and $\varepsilon_{\tau}^{R}$ (red line in right panel) are slightly deviated from zero, but, needless to mention that these deviations are statistically insignificant.

The one-dimensional log-likelihood profiles 
for both $\varepsilon_{\tau}^{R}$ and $\varepsilon_{\tau}^{L}$ look
non-parabolic in the top right panel of figure~\ref{fig:1d-profiles}.
In particular, $\varepsilon_{\tau}^{L}$ demonstrates one 
extra minimum around $\varepsilon_{\tau}^{L}$ $\approx$ 0.6,
which is slightly disfavored at $\Delta\chi^2 = - 2 \Delta ln L \approx 1.5$ as
compared to the global minimum at $\varepsilon_{\tau}^{L} = 0$.
This minimum originates due to the approximate 
$\tilde{g}_{\alpha L} \leftrightarrow - \tilde{g}_{\alpha L}$ symmetry 
that Eq.~\eqref{diffCS} possesses, since the first term in Eq.~\eqref{diffCS}
dominates over the third term~\cite{Bolanos:2008km}.
Because of this symmetry, the value of 
$\tilde{g}_{\tau L}^2 = (g_{\tau L}+\varepsilon_{\tau}^{L})^2$ 
is the same for $\varepsilon_{\tau}^{L} = 0$ and 
$\varepsilon_{\tau}^{L} = - 2 g_{\tau L} \approx 0.54$, 
and therefore, one may expect a local minimum in vicinity 
of the second point. The presence of the third term in 
Eq.~\eqref{diffCS} shifts the position of this local minimum 
slightly upward to $\varepsilon_{\tau}^{L} \approx 0.64$.

\begin{figure}[t]
\centering
\includegraphics[width=0.9\textwidth]{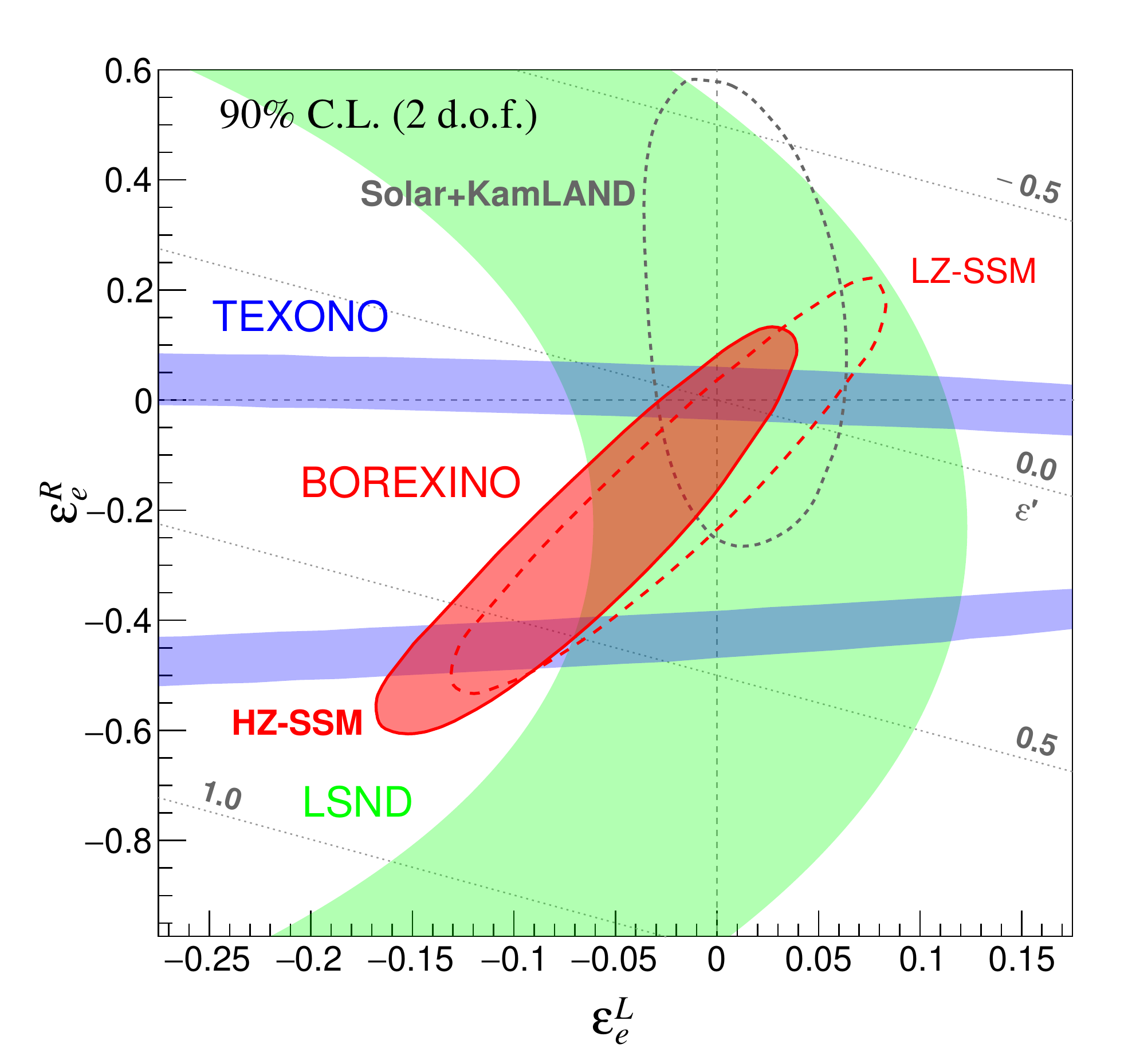}
\mycaption{Allowed region for NSI parameters 
in $\varepsilon_{e}^{L/R}$ plane obtained in the present work. 
The parameters $\varepsilon_{\tau}^{L}$ and $\varepsilon_{\tau}^{R}$ 
are fixed to zero. Both HZ- (filled red) and LZ- (dashed red) SSM's were assumed. The bounds from LSND \cite{Davidson:2003ha,Auerbach:2001wg} and 
TEXONO \cite{Deniz:2010mp} are provided for 
comparison. Besides, the contour obtained from the global analysis of solar neutrino experiments is presented by dashed black line (Ref. \cite{Bolanos:2008km}, NSI's are included in detection and propagation). All contours correspond to 90\% C.L. (2 d.o.f.). The dotted gray lines represent the corresponding range of $\varepsilon'$ parameter, relevant for NSI's at propagation.
}
\label{fig:eLeR2d}
\end{figure}

\begin{figure}[t]
\centering
\includegraphics[width=0.9\textwidth]{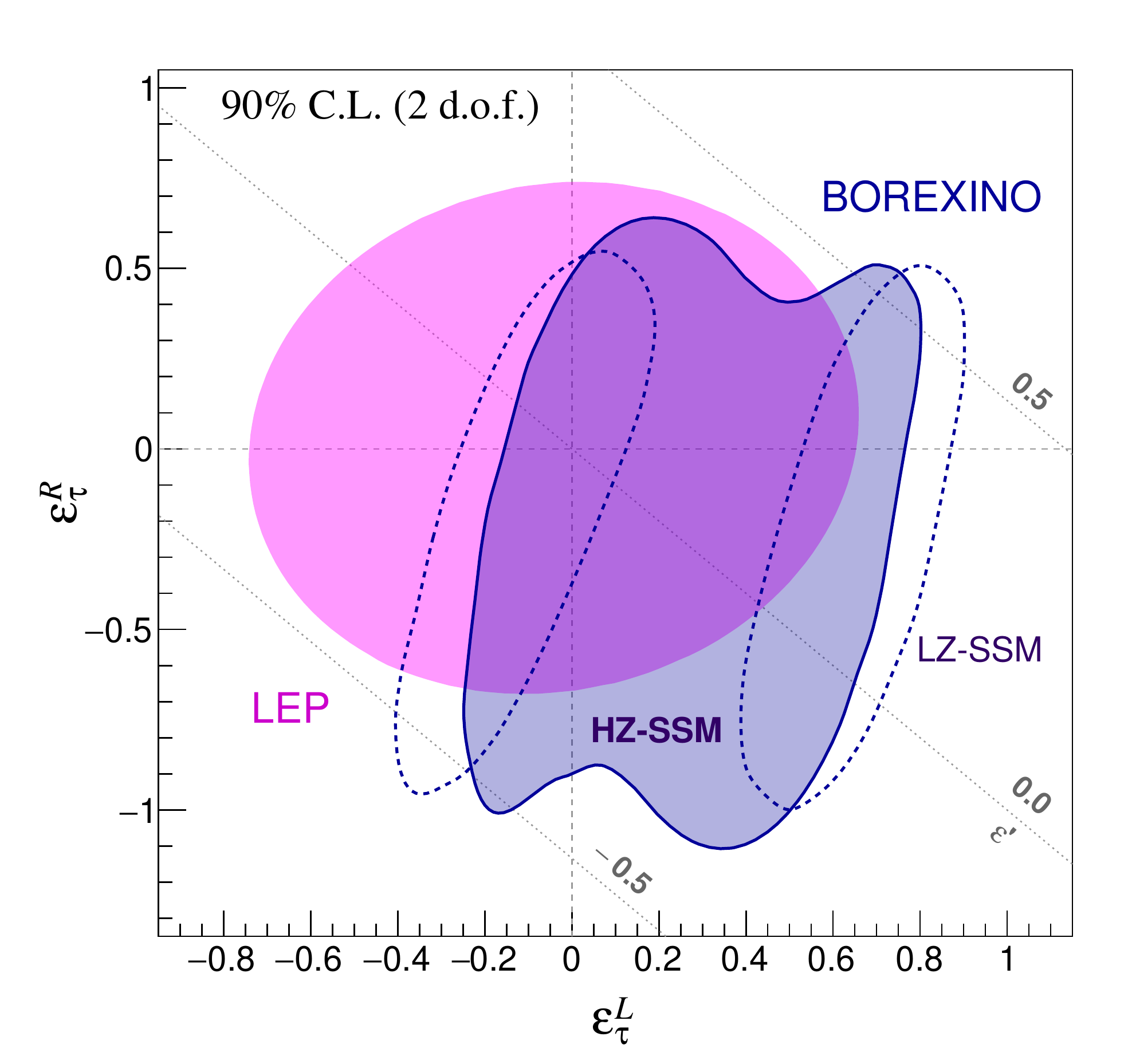}
\mycaption{Allowed region for NSI parameters in $\varepsilon_{\tau}^{L/R}$ plane obtained in the present work. The parameters $\varepsilon_{e}^{L}$ and $\varepsilon_{e}^{R}$ are fixed to zero. Both HZ- (filled dark blue) and LZ- (dashed dark blue) SSM's were assumed. The contour from LEP \cite{Barranco:2007ej} is provided for comparison. Both contours correspond to 90\% C.L. (2 d.o.f.). The dotted gray lines represent the corresponding range of $\varepsilon'$ parameter, relevant for NSI's at propagation.}
\label{fig:tauLtauR2d}
\end{figure}

The profiles for the LZ-SSM case (figure~\ref{fig:1d-profiles}, bottom panels) are clearly shifted from zero and with respect to the HZ-SSM ones. The main reason for this is that LZ-SSM predicts smaller  $\Phi_{{}^{7}\text{Be}}$ compared to HZ-SSM. The smaller flux requires a bigger cross section $\langle\sigma_{{}^{7}\text{Be}} \rangle$  for a given observed experimental rate $R_{{}^{7}\text{Be}}$ (see Eq. \ref{rate}).  As figure~\ref{fig:TotalCSeReL} illustrates, the total cross section linearly depends on $\varepsilon_{e}^{L}$. Therefore, the minimum for LZ-SSM should be shifted in positive direction of $\varepsilon_{e}^{L}$. For $\varepsilon_{\tau}^{L}$ the minima go in opposite directions due to the same reason. The only difference is that the cross section increases when $\varepsilon_{\tau}^{L}$ goes in negative direction for the first minimum and when $\varepsilon_{\tau}^{L}$ goes up for the second one (see figure~\ref{fig:TotalCSeReL}, right panel).
Aforementioned shifts for $\varepsilon_{e}^L$ and $\varepsilon_{\tau}^L$ profiles induce the shifts for $\varepsilon_{e}^R$ and $\varepsilon_{\tau}^R$ as well. This will be easy to see later on considering two dimensional profiles (figures~\ref{fig:eLeR2d} and \ref{fig:tauLtauR2d}).


\begin{table}[t]
\begin{center}

\begin{tabular}{|c||c|c||c||c|}
\hline 
& HZ-SSM & LZ-SSM & Ref.~\cite{Agarwalla:2012wf} & Ref.~\cite{Barranco:2007ej} \\
\hline\hline 
& & & &\\
$\varepsilon_{e}^{R}$   & [$-0.15$, $+0.11$~]   & [$-0.20$, $+0.03$~]  & [$-0.21$, $+0.16$~] & [$0.004$, $+0.151$~] \\
$\varepsilon_{e}^{L}$    & [$-0.035$, $+0.032$~] & [$-0.013$, $+0.052$~] & [$-0.046$, $+0.053$~] & [$-0.03$, $+0.08$~] \\
& & & &\\
$\varepsilon_{\tau}^{R}$ & [$-0.83$, $+0.36$~]   & [$-0.42$, $+0.43$~]   & [$-0.98$, $+0.73$~] & [$-0.3$, $+0.4$~] \\
$\varepsilon_{\tau}^{L}$ & [$-0.11$, $+0.67$~]   & [$-0.19$, $+0.79$~]   & [$-0.23$, $+0.87$~] & [$-0.5$, $+0.2$~] \\ 
& & & &\\
\hline
\end {tabular}


\end{center}

\mycaption{The first column shows the limits 
on the flavor-diagonal NSI parameters $\varepsilon_{e}^{R}$, 
$\varepsilon_{e}^{L}$, $\varepsilon_{\tau}^{R}$, and 
$\varepsilon_{\tau}^{L}$ as obtained in the present work
using the Borexino Phase-II data and considering HZ-SSM for 
the neutrino fluxes. The second column displays the same
considering LZ-SSM. These constraints are obtained varying
only one NSI parameter at-a-time, while the remaining three 
NSI parameters are fixed to zero. The third column contains 
the bounds using Borexino Phase-I results as obtained in 
Ref.~\cite{Agarwalla:2012wf} (for HZ-SSM case only). For the sake of comparison, 
we present the global bounds from Ref.~\cite{Barranco:2007ej} 
in the forth column. All limits are 90\% C.L. (1 d.o.f.).
}
\label{tab:be7pplimits}
\end{table}

The 90\% C.L. (1 d.o.f.) bounds on the flavor-diagonal
NSI parameters obtained using the Borexino Phase-II data
are listed in table~\ref{tab:be7pplimits}. The first column 
shows the constraints assuming HZ-SSM for the neutrino fluxes. The second column presents the same considering LZ-SSM. These constraints 
are obtained varying only one NSI parameter at-a-time, 
while the remaining three NSI parameters are fixed to zero.

The third column exhibits the bounds obtained by phenomenological analysis with Borexino Phase-I data in Ref.~\cite{Agarwalla:2012wf}. 
All experimental limits from Borexino Phase-II are better 
than those previously obtained from the Borexino Phase-I 
data in Ref.~\cite{Agarwalla:2012wf}. For the sake of 
comparison, in the forth column, we present the global bounds 
from Ref.~\cite{Barranco:2007ej}, where the authors analyzed
the data from the Large Electron-Positron Collider (LEP)
experiment, LSND and CHARM II accelerator experiments,
and Irvine, MUNU, and Rovno reactor experiments. The bounds found in the present analysis are quite comparable to the global ones. One may note that the best up-to-date bound for $\varepsilon_e^L$ was obtained in this work. \\

We have considered above the sensitivity of the Borexino Phase II data to NSI's applying the SSM-constraint on the neutrino fluxes. Remarkably, Borexino detector is sensitive to the modification of the shape of ${}^{7}\text{Be}$ electron recoil spectra even if the neutrino fluxes are not constrained by SSM model. Such analysis provides a limit: 
%
\begin{equation}
-1.14 < \epsilon_e^{R} < 0.10 \quad ( 90\% \: \text{C.L.}. )
\end{equation} 
%
As one may see the limit is highly asymmetric, with a large extension for the negative values of $\varepsilon_e^{R}$. Such a small sensitivity is induced by backgrounds (mostly ${}^{85}\text{Kr}$) which can easily compensate the modification of electron-recoil spectra. \\





%

Now let us consider the two-dimensional case when the allowed region for NSI parameters $\varepsilon_{e}^{L/R}$ is plotted while $\varepsilon_{\tau}^{L}$ and $\varepsilon_{\tau}^{R}$ are fixed to zero (figure~\ref{fig:eLeR2d}). 
Two contours for HZ- (filled) and LZ-SSM (dashed) were obtained. 
Compared with other experiments sensitive to the same NSI's, the allowed contours for Borexino in the $\varepsilon_{e}^{L}$-$\varepsilon_{e}^{R}$ plane have a distinct orientation, \textit{cf.} figure~\ref{fig:eLeR2d}.  
The TEXONO experiment \cite{Deniz:2010mp} is mostly sensitive to $\varepsilon_{e}^{R}$,
while LSND \cite{Davidson:2003ha,Auerbach:2001wg} is mostly sensitive to $\varepsilon_{e}^{L}$ . 
Borexino's contour intersects the allowed regions for both experiments at a certain angle, and the three experiments complement each other. 
In principle, the overlap of Borexino with TEXONO results in two allowed regions. To exclude the second intersection, the incorporation of the LSND result is necessary.

As it was already explained in the analysis of one-dimensional profiles, the contours for HZ- and LZ-SSM's are shifted along $\varepsilon_e^L$-axis. Considering two-dimensional case it is evident that such a shift in $\varepsilon_e^L$ has to produce also the displacement for $\varepsilon_e^R$.

The contour for Borexino is extended in the direction of negative $\varepsilon_e^R$ and $\varepsilon_e^L$ due to the presence of backgrounds, especially ${}^{85}\text{Kr}$. The shift of the contour for LZ-case with respect to HZ one is due to the change of rate of the backgrounds because of the spectral correlations.

%
%


For both HZ- and LZ-SSM cases, the bounds on the left parameter are stronger than the result from LSND.  
TEXONO \cite{Deniz:2010mp} is a reactor antineutrino experiment and its bounds are obtained from $\overline{\nu}_e e$ scattering.
For anti-neutrinos the roles of $\tilde{g}_{eL}$ and $\tilde{g}_{eR}$ are reversed, leading to
a stronger bound on $\varepsilon_{e}^{R}$.
Due to the approximate symmetry $\tilde{g}_e^R \leftrightarrow -\tilde{g}_e^R$ in the anti-neutrino scattering cross section, two separate contours form the allowed region of TEXONO around $\varepsilon_e^R = 0$ and $\varepsilon_e^R = -2g_{eR} = -2 \sin^2\theta_W \approx -0.5$.

The contour obtained in Ref. \cite{Bolanos:2008km} is presented by a dashed black line. In this work, the global analysis of several solar neutrino experiments together with KamLAND result was conducted. NSI's were considered in both detection and propagation. The very first results of Borexino were also included \cite{collaboration2007first,Bellini:2008mr,Arpesella:2008mt}. Though, as the authors underlined, they did not contribute much in overall sensitivity to NSI's. As one may see, the present Borexino results are quite complementary to this contour.  


The result of Borexino in the $\varepsilon_{\tau}^{L}$-$\varepsilon_{\tau}^{R}$ 
plane is shown in figure~\ref{fig:tauLtauR2d}. It is similar to that of LEP \cite{Barranco:2007ej} in excluded area, but it occupies a slightly different region, 
favoring positive $\varepsilon_{\tau}^{R}$ and negative $\varepsilon_{\tau}^{L}$. 
%
NSI's comparable with the SM neutral current interactions are still allowed.

The result for LZ-SSM is of particular interest. The shift of the minima discussed above and observed in figure~\ref{fig:1d-profiles} (bottom right) transforms the allowed contour (figure~\ref{fig:tauLtauR2d}, dashed dark blue) into two separate regions, one of which is already almost completely excluded by LEP data. So, the remaining allowed region, in this case, is relatively small.


The dotted gray lines in figures \ref{fig:eLeR2d} and \ref{fig:tauLtauR2d}  indicate the range for the parameter $\varepsilon'$ relevant for NSI's at propagation. The contours are almost entirely located between $\varepsilon'=-0.5$ and $\varepsilon'=0.5$. As it was previously shown in section~\ref{NSIsection} (see figure~\ref{fig:PeeNSI}),  NSI's at propagation are not very pronounced for these magnitudes of $\varepsilon'$ compare to the precision of the the measurements. Thus, the sensitivity of the detector is mostly determined by NSI's at detection.

\subsection{Evaluation of $\sin^2\theta_W$}

In addition to the analysis of NSI's, 
we use the same data and analysis approach to constrain the value of $\sin^2\theta_W$.
Instead of introducing NSI's, we simply allow $\sin^2\theta_W$ in the SM couplings \eqref{gLgR} to vary. The sensitivity of the analysis to $\sin^2 \theta_W$ is mostly dominated by $g_e^L$, while contributions of the other five coupling constants are almost negligible.
For the HZ-SSM case, the analysis of a likelihood profile results in
\begin{equation}
\sin^2\theta_W \;=\; 0.229 \pm 0.026 \: \text{(stat+syst)}\;, 
\end{equation} 
which is consistent with theoretical expectations \cite{Erler:2018} and comparable in precision with 
the value found by the reactor $\overline{\nu}_e e$ scattering experiment TEXONO \cite{Deniz:2010mp}:
\begin{equation}
\sin^2 \theta_W \;=\; 0.251 \pm 0.031 \: (\text{stat}) \pm 0.024 \: (\text{syst})\;.
\end{equation}
The most accurate determination of $\sin^2\theta_W$ by neutrino-electron scattering is from
the $\nu_\mu e$ scattering experiment CHARM II \cite{Vilain:1994qy}:
\begin{equation}
\sin^2 \theta_W \;=\; 0.2324 \pm \: 0.0058 \: (\text{stat}) \pm \: 0.0059 \: (\text{syst})\;.
\end{equation}



\section{Summary and Concluding Remarks}
\label{sec:conclusions}

In the present work, we search for Non-Standard Interactions (NSI's) of the neutrino using Borexino Phase-II data. 
The NSI's considered are those of the flavor-diagonal neutral current type that modify the $\nu_e e$ and $\nu_\tau e$ couplings while preserving their chiral and flavor structures.

Such NSI's can affect solar neutrinos at production, propagation, and detection.
Neutrino production in the Sun can be affected via processes such as $\gamma e\to\nu\bar{\nu}e$, but the expected modification 
in the neutrino spectrum is at energies well below the detection threshold of Borexino ($\sim$50 keV), so this effect does not need to be included.
The NSI's considered also modify the solar neutrino survival probability $P_{ee}(E)$ via the LMA-MSW effect
as the neutrinos propagate through dense solar matter.  This effect is strong at ${}^{8}\text{B}$ neutrino energies
but not particularly large at ${}^{7}\text{Be}$ neutrino energies, limiting the sensitivity of Borexino to such deviations.
The effect of the NSI's to which Borexino is most sensitive is at detection, where the shape of the electron-recoil spectrum is affected by changes in the $\nu_e e$ and $\nu_\tau e$ couplings.

The solar neutrino fluxes were constrained to the prediction of the Standard Solar Model (SSM) with the LMA-MSW oscillation mechanism. 
SSM's with both high- (HZ) and low-metallicity (LZ) were considered. Systematic effects related to the characterization of the target mass of the detector and the choice of oscillation parameters were taken into account. 

The modifications to the $\nu_e e$ and $\nu_{\tau} e$ couplings are quantified by parameters 
$\varepsilon_e^{L/R}$ and $\varepsilon_\tau^{L/R}$.
The bounds to all four parameters were obtained in this analysis, and they all
show marked improvement compared to the Borexino Phase-I analysis \cite{Agarwalla:2012wf},
regardless of the choice of metallicity in the SSM, \textit{cf.} table~\ref{tab:be7pplimits}.
The bounds are quite comparable to the global ones. 
In particular, the best constraint to-date on $\varepsilon_e^L$ was obtained.

The log-likelihood profiles and corresponding bounds for HZ- and LZ-SSM's are shifted with respect to each other due to different expected neutrino detection rates. The minima of HZ-profiles are less shifted from zero as a result of better agreement between the measured neutrino rates and HZ-SSM. 
For LZ-SSM, the deviations of the minima of the profiles from zero are more pronounced but still statistically insignificant. 
The allowed contour of Borexino in the $\varepsilon_{e}^{L/R}$-plane is quite distinct with respect to other $\nu e$ or $\overline{\nu} e$ scattering experiments, also sensitive to the same NSI's, such as TEXONO and LSND. 
Borexino is sensitive to both $\varepsilon_{e}^{R}$ and $\varepsilon_{e}^{L}$ parameters while TEXONO and LSND mostly constrain $\varepsilon_{e}^{R}$ or $\varepsilon_{e}^{L}$, respectively.
Notably, in the case of $\varepsilon_{\tau}^{L/R}$ two local minima are observed. The distance between the minima is larger for LZ-SSM, resulting in the splitting of the 90\% C.L. allowed contour into two contours in the $\varepsilon_{\tau}^{L/R}$-plane. 

An important sensitivity-limiting factor is the presence of backgrounds, especially ${}^{85}\text{Kr}$, whose forbidden $\beta$-spectrum can mimic the spectral modifications induced by NSI's. 

The smaller, conservative, NH-value for $\theta_{23}$ was chosen for the $\varepsilon_{\tau}^{L/R}$-analysis. 
Should the neutrino mass hierarchy be identified as inverted in future experiments, the contribution of the $\tau$-neutrino to the cross section would be larger and the bounds for $\varepsilon_{\tau}^{L/R}$ would be slightly improved. 
The most important factor which determines the sensitivity of this study is the uncertainty on the $\nu$-fluxes predicted by the SSM (currently 6\% for $\Phi_{{}^{7}\text{Be}}$). Their improvement would directly refine the bounds on NSI's presented here.

The detector is sensitive to $\varepsilon_e^{R}$, even without constraining the solar neutrino fluxes to those of the SSM,
purely via the modification to the electron-recoil spectral shape. 
However, it was found that background greatly reduce the ideal sensitivity by compensating for the modification to the spectra, especially for negative $\varepsilon_e^{R}$. 

The same dataset and approach, but without any NSI's assumed, 
was used to constrain $\sin^2\theta_W$. 
The resulting value is comparable in precision to that measured in reactor antineutrino experiments.

\section*{Acknowledgments}

The Borexino program is made possible by funding from INFN (Italy), 
NSF (USA), BMBF, DFG, HGF and MPG (Germany), RFBR 
(grants 19-02-0097A, 16-29-13014ofi-m, 17-02-00305A, 16-02-01026), RSF (grant 17-12-01009) 
(Russia), and NCN (grant number UMO 2017/26/M/ST 2/00915) (Poland). 

We acknowledge also the computing services of the Bologna INFN-CNAF 
data centre and LNGS Computing and Network Service (Italy), of J\"{u}lich 
Supercomputing Centre at FZJ (Germany), of ACK Cyfronet 
AGH Cracow (Poland), and of HybriLIT (Russia).
We acknowledge the hospitality and support of the 
Laboratori Nazionali del Gran Sasso (Italy).

We are thankful to Zurab Berezhiani for the discussion at the initial stage of the research. S.K.A. would like to thank A.~Smirnov and A.~De~Gouvea for useful discussions. S.K.A. acknowledges the support from DST/INSPIRE Research Grant [IFA-PH-12], Department of Science and Technology, India 
and the Young Scientist Project [INSA/SP/YSP/144/2017/1578] from the Indian National Science Academy.  A.F. acknowledges the support provided by the University of Hamburg. T.T. is supported in part by NSF Grant 1413031. C.S. is supported in part by the International Postdoctoral Fellowship funded by China Postdoctoral Science Foundation, and NASA grant 80NSSC18K1010.


\newpage
\appendix
\section{Derivation of the Matter Effect Potential in the presence of NSI's}\label{Vprime}

The Hamiltonian which governs the propagation of neutrinos in matter 
in the presence of the NSI's 
$\varepsilon_{e}^{V}=\varepsilon_{e}^{L}+\varepsilon_{e}^{R}$ and 
$\varepsilon_{\tau}^{V}=\varepsilon_{\tau}^{L}+\varepsilon_{\tau}^{R}$ 
is given by
\begin{equation}
H \;=\; \dfrac{1}{2E}\;U 
\begin{bmatrix} 0 & 0 & 0 \\ 0 & \Delta m_{21}^2 & 0 \\ 0 & 0 & \Delta m_{31}^2 \end{bmatrix}
U^\dagger
+V(x)
\begin{bmatrix} 1+\varepsilon_{e}^{V} & \ 0 & \ 0 \ \\ 0 & \ 0 & \ 0 \ \\ 0 & \ 0 & \ \varepsilon_{\tau}^{V} \end{bmatrix}\;,
\label{Hzero}
\end{equation}
where $V(x)=\sqrt{2}G_F N_e(x)$, and $N_e(x)$ is the electron density at location $x$.
This expression is in the flavor basis in which the rows and columns are labelled by neutrino-flavor in the order $(e,\mu,\tau)$. 
The Pontecorvo-Maki-Nakagawa-Sakata (PMNS) matrix $U$ is given by
\begin{eqnarray}
U 
& = &
\underbrace{
\left[ \begin{array}{ccc} 1 & 0 & 0 \\
                          0 &  c_{23} & s_{23} \\
                          0 & -s_{23} & c_{23}
       \end{array}
\right]}_{\displaystyle R_{23}}
\underbrace{
\left[ \begin{array}{ccc} c_{13} & 0 & s_{13} e^{-i\delta} \\
                          0 & 1 & 0 \\
                          -s_{13} e^{i\delta} & 0 & c_{13}
       \end{array}
\right]}_{\displaystyle R_{13}}
\underbrace{
\left[ \begin{array}{ccc} c_{12} & s_{12} & 0 \\
                         -s_{12} & c_{12} & 0 \\
                          0 & 0 & 1
       \end{array}
\right]}_{\displaystyle R_{12}}
\;.
%
\label{UPparam}
\end{eqnarray}
Performing the $R_{23}$ and $R_{13}$ rotations on both sides of Eq.~\eqref{Hzero}, we find
\begin{eqnarray}
H' 
& = & R_{13}^\dagger R_{23}^\dagger H R_{23} R_{13}
\cr
& = & \dfrac{1}{2E}\;R_{12}
\begin{bmatrix} 0 & 0 & 0 \\ 0 & \Delta m_{21}^2 & 0 \\ 0 & 0 & \Delta m_{31}^2 \end{bmatrix}
\!R_{12}^\dagger  
+V(x)\,
R_{13}^\dagger R_{23}^\dagger
\begin{bmatrix} 1+\varepsilon_{e}^{V} & \ 0 & \ 0 \ \\ 0 & \ 0 & \ 0 \ \\ 0 & \ 0 & \ \varepsilon_{\tau}^{V} \end{bmatrix}
\!R_{23} R_{13}
\cr
& = &
\dfrac{1}{2E}\;
\begin{bmatrix} 
\Delta m_{21}^2 s_{12}^2 & \Delta m_{21}^2 s_{12}c_{12} & 0 \\ 
\Delta m_{21}^2 s_{12}c_{12} & \Delta m_{21}^2 c_{12}^2 & 0 \\ 
0 & 0 & \Delta m_{31}^2 
\end{bmatrix} 
\cr
& &
+V(x)
\begin{bmatrix} 
(1-\varepsilon')c_{13}^2 + \varepsilon_{\tau}^{V}(c_{23}^2 - s_{23}^2) s_{13}^2 
& \ \varepsilon_{\tau}^{V}s_{23}c_{23}s_{13}e^{-i\delta} 
& \ \left\{(1+\varepsilon_{e}^{V})-\epsilon_{\tau}^{V}c_{23}^2\right\}s_{13}c_{13}e^{-i\delta} \ 
\\ 
\varepsilon_{\tau}^{V}s_{23}c_{23}s_{13}e^{i\delta} 
& 0 
& \ -\varepsilon_{\tau}^{V}s_{23}c_{23}c_{13} \ 
\\ 
\left\{(1+\varepsilon_{e}^{V})-\epsilon_{\tau}^{V}c_{23}^2\right\}s_{13}c_{13}e^{i\delta} \ 
& \ -\varepsilon_{\tau}^{V}s_{23}c_{23}c_{13} \ 
& \ \varepsilon_{\tau}^{V}(c_{23}^2 - s_{23}^2)c_{13}^2 + (1-\varepsilon')s_{13}^2 
\end{bmatrix}
\cr
& & 
+ \;\;V(x)\,\varepsilon_\tau^V s_{23}^2\times\mbox{(unit matrix)}\;,\vphantom{\Bigg|}
\end{eqnarray}
where we have set
\begin{equation}
\varepsilon' \;=\; \varepsilon_{\tau}^{V} s_{23}^2 - \varepsilon_{e}^{V}\;.
\end{equation}
In the energy range where $|\Delta m^2_{31}| \gg 2EV(x)$,
the off-diagonal terms in the third row and third column can be neglected and  
we can treat $H'$ as already partially diagonalized.
Concentrating on the $2\times 2$ upper-left block, we drop the third row and third column and
obtain
\begin{eqnarray}
H' & \to &
\dfrac{1}{2E}\;
\begin{bmatrix} 
\Delta m_{21}^2 s_{12}^2 & \Delta m_{21}^2 s_{12}c_{12} \\ 
\Delta m_{21}^2 s_{12}c_{12} & \Delta m_{21}^2 c_{12}^2 \\ 
\end{bmatrix} 
\cr
& & 
+V(x)
\begin{bmatrix} 
(1-\varepsilon')c_{13}^2 + \varepsilon_{\tau}^{V}(c_{23}^2 - s_{23}^2) s_{13}^2 
& \ \varepsilon_{\tau}^{V}s_{23}c_{23}s_{13}e^{-i\delta} 
\\ 
\varepsilon_{\tau}^{V}s_{23}c_{23}s_{13}e^{i\delta} 
& 0 
\end{bmatrix}
\cr
& \approx &
\dfrac{1}{2E}\;
\begin{bmatrix} 
\Delta m_{21}^2 s_{12}^2 & \Delta m_{21}^2 s_{12}c_{12} \\ 
\Delta m_{21}^2 s_{12}c_{12} & \Delta m_{21}^2 c_{12}^2 \\ 
\end{bmatrix} 
+V(x)
\begin{bmatrix} 
(1-\varepsilon')c_{13}^2 
& \ \varepsilon_{\tau}^{V}s_{23}c_{23}s_{13}e^{-i\delta} 
\\ 
\varepsilon_{\tau}^{V}s_{23}c_{23}s_{13}e^{i\delta} 
& 0 
\end{bmatrix}
\;.
\cr & & 
\end{eqnarray}
where we have used $s_{23}^2\approx c_{23}^2$ and $s_{13}^2\ll 1$ to simplify the expression for the $(1,1)$ element.
The resonance condition is achieved when the $(1,1)$ and $(2,2)$ elements of this matrix are equal:
\begin{equation}
\Delta m^2_{21}s_{12}^2 + 2EV'(x)\,c_{13}^2 \;=\; \Delta m^2_{21}c_{12}^2
\quad\to\quad
2EV'(x)\,c_{13}^2 \;=\; \Delta m^2_{21}\cos 2\theta_{12}\;.
\end{equation}
where
\begin{equation}
V'(x) \;=\; V(x)\,(1-\varepsilon')\;.
\end{equation}
See, e.g., Ref.~\cite{Agarwalla:2013tza, Agarwalla:2015cta} and references therein.

\newpage

\bibliographystyle{JHEP}
\bibliography{NSIfinal}

\end{document}